\documentclass[12pt]{article}
\usepackage[utf8]{inputenc}
\usepackage[usenames,dvipsnames]{color}
\usepackage{amsmath}
\usepackage{amssymb}
\usepackage{mathptmx}
\usepackage{authblk}

\usepackage[a4paper, margin=1in]{geometry}
\usepackage[left]{lineno}

\usepackage{multirow}
\usepackage{array}
\usepackage{rotating}
\usepackage{booktabs}%

\usepackage[breaklinks,colorlinks,citecolor=RoyalBlue,linkcolor=magenta]{hyperref}

\usepackage[super,compress,comma,sectionbib]{natbib}
\usepackage{graphicx}
\usepackage{subcaption}
\usepackage[labelfont=bf]{caption}
\DeclareCaptionLabelSeparator{bar}{ \textbar{} }
\newcommand{\arcsec}{^{\prime\prime}}
\newcommand{\arcmin}{^{\prime}}
\newcommand{\arcdeg}{^{\circ}}
\newcommand{\degr}{^{\circ}}

\def\jnl@style{\it}
\def\aaref@jnl#1{{\jnl@style#1}}

\newcommand{\araa}{Annu. Rev. Astron. Astrophys.}   
\newcommand{\aj}{Astron. J.}   
\newcommand{\apj}{Astrophys. J.}   
\newcommand{\apjl}{Astrophys. J. Lett.}   
\newcommand{\apjs}{Astrophys. J. Suppl. Ser.}   
\newcommand{\aap}{Astron. Astrophys.}   
\newcommand{\aaps}{Astron. Astrophys. Suppl.}   
\newcommand{\mnras}{Mon. Not. R. Astron. Soc.}   
\newcommand{\nat}{Nature} 

\newcommand{\HI}{H{\sc i}}
\newcommand{\sun}{_\odot}
\usepackage{etoolbox}
\makeatletter
\patchcmd{\@maketitle}{\null\vskip 2em}{}{}{}
\makeatother

\newenvironment{myabstract}{%
  \list{}{%
    \leftmargin1.5em
    \rightmargin0pt}%
    \item\relax
    \small}
{\endlist} 

\title{Network of velocity-coherent filaments formed by supersonic turbulence in a very-high-velocity HI cloud}  

\author[1*]{Xunchuan Liu}%
\affil[1]{Shanghai Astronomical Observatory, Chinese Academy of Sciences, Shanghai 200030, PR China; 
\url{liuxunchuan@shao.ac.cn}; \url{liutie@shao.ac.cn}; \url{zshen@shao.ac.cn}}%

\author[1*]{Tie Liu}%

\author[1]{Pak-Shing Li}%

\author[1]{Xiaofeng Mai}

\author[2]{Christian  Henkel}
\affil[2]{
Max-Planck-Institut f\"ur Radioastronomie, Auf dem H\"ugel 69, 53121 Bonn, Germany}

\author[3]{Paul F. Goldsmith}
\affil[3]{Jet Propulsion Laboratory, California Institute of Technology, 4800 Oak Grove Drive, Pasadena CA 91109, USA}

\author[4]{Sheng-Li Qin}
\affil[4]{School of Physics and Astronomy, Yunnan University, Kunming, 650091, PR China}

\author[5]{Yan Gong}
\affil[5]{Purple Mountain Observatory, Chinese Academy of Sciences, 10 Yuanhua Road, Nanjing 210033, PR China}

\author[1]{Xing Lu}

\author[6,7]{Fengwei Xu}
\affil[6]{Kavli Institute for Astronomy and Astrophysics, Peking University, Haidian District, Beijing 100871, PR China}
\affil[7]{Department of Astronomy, School of Physics, Peking University, Beijing 100871, PR China}

\author[1]{Qiuyi Luo}

\author[4]{Hong-Li Liu}

\author[8]{Tianwei Zhang}
\affil[8]{Research Center for Astronomical computing, Zhejiang Laboratory, Hangzhou, PR China}

\author[9]{Yu Cheng}
\affil[9]{National Astronomical Observatory of Japan, 2-21-1 Osawa, Mitaka, Tokyo, 181-8588, Japan}

\author[10,1]{Yihuan Di}
\affil[10]{School of Physics and Astronomy, Shanghai Jiao Tong University, Shanghai 200240, PR China}

\author[7]{Yuefang Wu}

\author[1]{Qilao Gu}

\author[11]{Ningyu Tang}
\affil[11]{Department of Physics, Anhui Normal University, Wuhu, Anhui 241002, PR China}

\author[12]{Aiyuan Yang}
\affil[12]{National Astronomical Observatories, Chinese Academy of Sciences, Beijing 100101, PR China}

\author[1*]{Zhiqiang Shen}
\begin{document}

\maketitle             %

\begin{myabstract}

The warm neutral medium (WNM)  was thought to be subsonically/transonically turbulent, and it lacks a network of intertwined filaments  that are commonly seen in both molecular clouds and cold neutral medium (CNM).
Here, we report \HI~21 cm line observations of a very-high-velocity (-330 km s$^{-1}$ $<V_{\rm LSR}<$ -250 km s$^{-1}$) cloud (VHVC), using the Five-hundred-meter Aperture Spherical radio Telescope (FAST) with unprecedented resolution and sensitivity.
For the first time, such a VHVC is clearly revealed to be a supersonic WNM system consisting of a network  of velocity-coherent \HI~filaments.
The filaments are in the forms of slim curves, hubs, and webs, distributed 
in different layers within the position-position-velocity ({\it ppv}) data cube.
The entire cloud has skewed log-normal probability distribution of column density and the filaments themselves show asymmetrical radial density profiles,
indicating shock compression by supersonic magnetohydrodynamic (MHD) turbulence, as is also confirmed by our MHD simulation (sonic Mach number $M_{\rm s}=3$ and Alfv\'en Mach number $M_{\rm A}=1$).
This work suggests that hierarchical filaments can be established by shocks in a low-density WNM, where gravity is negligible, offering a viable pathway to structure formation in the earliest evolutionary phases of the interstellar medium (ISM).

\end{myabstract}

\phantomsection
\label{sec:Main body}
\noindent{\large \textbf{Main body}}

\begin{figure*}[!h]
\centering
\includegraphics[width=0.999\linewidth]{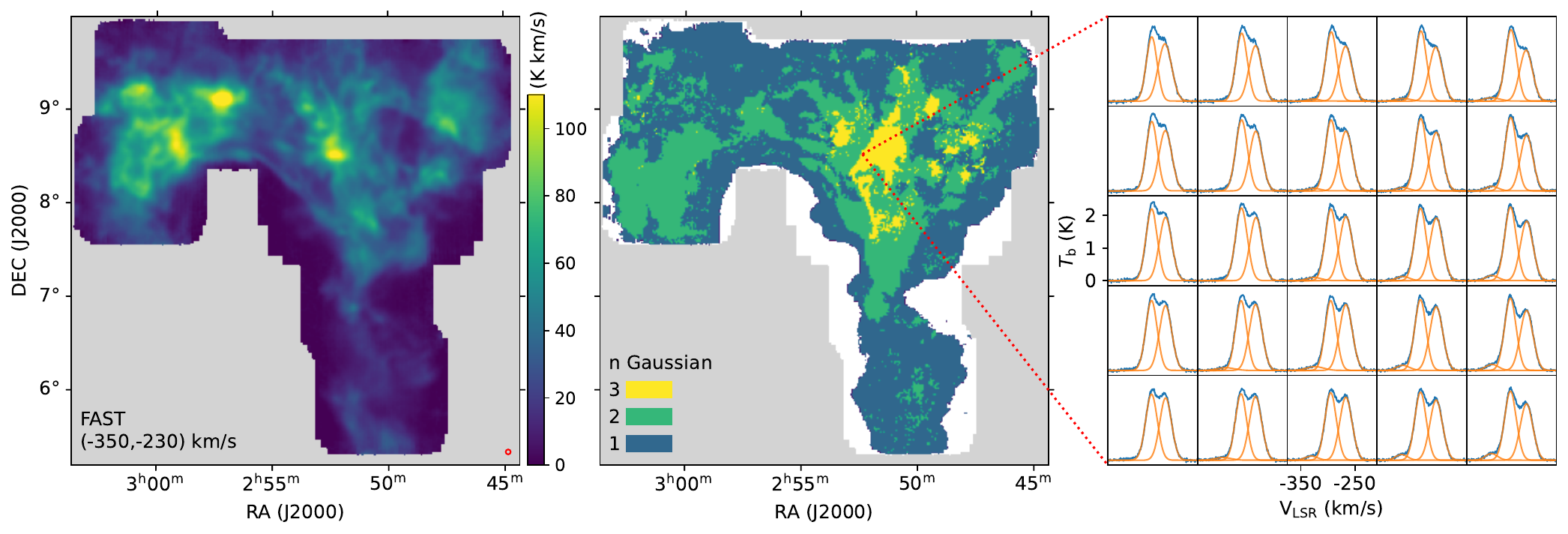}
\caption{
Spatial distribution and example spectra of the H{\sc i} 21\,cm lines in the G165 VHVC.
\textbf{Left panel}: The integrated intensity map of the \HI~21 cm emission of the G165 VHVC by FAST
(Sect. Methods). Red circle in the lower right corner
denotes the FAST beam size. \textbf{Middle panel}: The number of 
\HI~Gaussian subcomponents of the G165 VHVC at each pixel decomposed using {\it scousepy} (Sect. Methods). \textbf{Right panel}:
Example of the \HI~spectra (blue) and Gaussian decomposition (orange).
The residuals of the Gaussian fitting are presented in Supplementary Fig. 5.
\label{fig_quicklook}  
}
\end{figure*}

Hierarchical density structures in the warm neutral medium (WNM) are not yet well characterized by observations \citep{2009ARA&A..47...27K, 2023ARA&A..61...19M}, in contrast to the clearer picture emerging for molecular clouds \citep{2010A&A...518L.102A,2015A&A...584A..91K,2016A&A...591A...5L,2018ApJS..234...28L,2021ApJ...912..148L,2022MNRAS.514.6038Z,2023ASPC..534..153H,2024RAA....24b5009L} and the cold neutral medium (CNM) \citep{2014ApJ...789...82C,2016ApJ...821..117K,2023MNRAS.526.4345K}, even though theory predicts that sub-Alfvénic turbulence can give rise to elongated, low-density filaments in the WNM \citep{XuSY_2019}. 
High-velocity clouds (HVCs; $|V_{\rm LSR}|>100$ km s$^{-1}$) that contain a significant amount of WNM gas and locate in relatively isolated environments provide an excellent laboratory to explore this issue.
HVCs are \HI~clouds at radial velocities incompatible with the regular rotation of the 
Galactic disc\citep{1963CRAS..257.1661M,1991A&A...250..499W}, and the absorption lines of distant halo stars
indicate that they are generally located several kpc above the Galactic plane
(refs.\citep{1999Natur.400..138V,2008ApJ...672..298W,2008ApJ...684..364T,2016ApJ...828L..20P}).
The overall negative velocities of HVCs (refs.\citep{1991A&A...250..499W,2008ApJ...672..298W,2018MNRAS.474..289W})
suggest that they may trace the infall  of early-phase low-metalicity material onto the plane of the Milky Way
(refs.\citep{1980ApJ...236..577B,2004A&A...419..527D,2016ApJ...816L..11F,2023ApJ...944...65C}).
High-Velocity Clouds (HVCs), as observed in all-sky \HI~surveys such as HI4PI \citep{2018MNRAS.474..289W}, typically appear as atomic gas streams or complexes. This is likely due to the limited resolution of these surveys (~16$\arcmin$; see also Extended Data Fig. 1). A few HVCs with $|V_{\rm LSR}|<200$ km/s have been observed by single-dish telescopes such as Arecibo \citep{Peek_2018} and GBT \citep{2015ApJ...809..153M}, as well as interferometers \citep{2017ApJ...834..126B}. These observations have revealed that HVCs are primarily filamentary in nature \citep{2008ApJ...679L..21L, 2024AAS...24340224H, 2025MNRAS.536.3507H}, with many filaments aligned in parallel patterns. However, interferometric observations have revealed much more complex filamentary structures within HVCs, and narrow-line cold neutral medium (CNM) components have been detected in these \HI~filamentary systems \citep{2021ApJ...921...11M,2023ApJ...951..120V}. It is possible that the strip-shaped filamentary structures observed in HVCs are influenced by the global dynamics of the gas streams or by dynamic processes within the Galactic plane \citep{2024AAS...24340224H, 1991A&A...250..509W}, while the transition from warm neutral medium (WNM) to CNM contributes to the formation of more complex filamentary structures \citep{2021ApJ...921...11M,2023ApJ...951..120V}. In contrast, very-high-velocity clouds (VHVCs) are not coupled with the Galactic disc and are thought to have a different origin from typical HVCs \citep{1991A&A...250..509W,1981AJ.....86.1468G,1978A&A....66L...5H,1980AJ.....85.1155G}. 
Among them, VHVCs with $|V_{\rm LSR}| > 200$ km/s and located offset from the Galactic plane are clearly not co-rotating with the Galactic disk and, therefore, serve as the fiducial sample of VHVCs.
Those VHVCs may represent an earlier evolutionary stage compared to HVCs and could still primarily contain WNM. Exploring the detailed inner structures of VHVCs, particularly their possible filamentary patterns, is essential for understanding the early formation of density structures in atomic clouds. However, detailed studies of VHVCs have been extremely rare in the past. Observations of the \HI~21 cm line ($^{2}S_{1/2}$, $F$=1-0; $f_0=1420.4058$ MHz) \citep{1963CRAS..257.1661M}, with sufficiently high resolution to reveal the internal morphology and gas dynamics of VHVCs, are crucial for studying their origins and understanding the physical processes involved in the formation of hierarchical gas structures within the WNM.


\begin{figure*}[!h]
\centering
\includegraphics[width=0.98\linewidth]{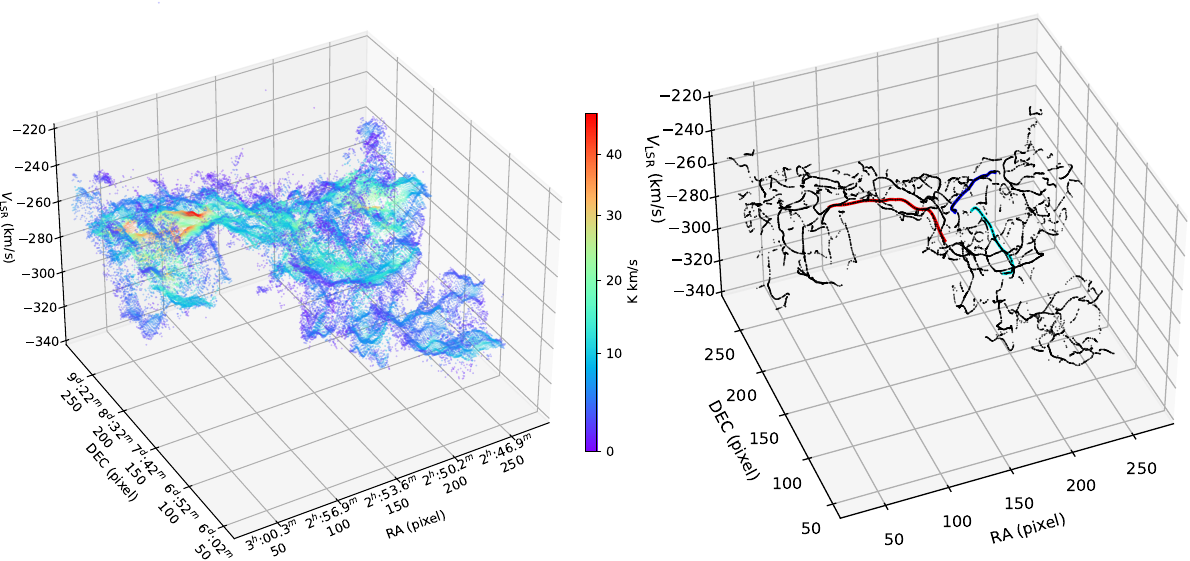}
\caption{H{\sc i} distribution of G165 VHVC in {\it ppv} space.
\textbf{Left panel}: The distribution of the decomposed Gaussian components of G165 VHVC. Each point represents the location (RA and DEC)
and central velocity ($V_{\rm LSR}$) of the corresponding velocity component. The pixel size is 1$\arcmin$. 
The color represents the integrated intensity of the
Gaussian components. 
\textbf{Right panel}: The \HI~filament ridges of VHVC G165 in the {\it ppv} space.
Three typical filaments, B1, B2, and B3, are represented by different colors (see also Extended Data Fig. 4). 
\label{fig_ppvdots}} 
\end{figure*}

\begin{figure*}[h]
\centering
\makebox[\textwidth][c]{
\includegraphics[width=1.2\textwidth]{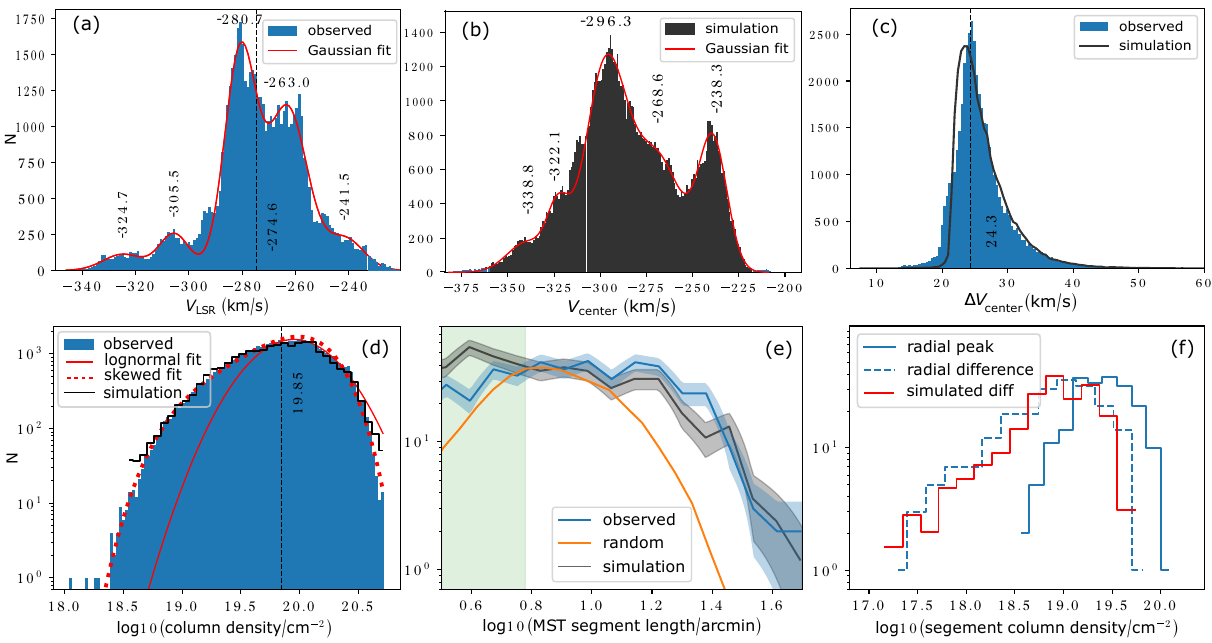}
}
\caption{Statistical properties of the H{\sc i} layers and filaments in the G165 VHVC.
\textbf{a-b}: The probability distribution functions (PDFs) of the decomposed velocity of the subcomponents in the observed and best-fit simulated data ($M_{\rm s}=3$ and $M_{\rm A}=1$). The red line shows the multi-Gaussian fitting. The
velocity has been shifted to roughly match the velocity range of observation. 
\textbf(c): The PDFs of the decomposed FWHM line widths for the observations
(blue area) and simulation (black solid line).
The black dashed line denotes the line width at the peak of the black area. 
\textbf(d) The PDFs of the decomposed column density for the observations
(blue area) and the best-fit simulation (black line).
The red solid line is the lognormal fit to the high-end values of the
observation  ($\log_{10}(N)>19.85$), and the 
red-dotted line shows the fourth order polynomial fitting of log(y). 
See Extended Data Fig. 8 for other models.
\textbf{e}: The PDFs of the filament-segment length for observation (Extended Data Fig. 5) and best-fit simulation (Extended Data Fig. 7),
and random distribution (Sect Methods). The gray and blue shadows represent 1$\sigma$ errors. The green shadow delineates scales smaller than 6$\arcmin$ that can not be well resolved. 
\textbf{f}: PDFs of the column density of observed filament segments (solid)
and the column density difference between the two sides of the filament segments (dashed). The red line is the simulated PDF of the column density difference, which has been normalized for better comparison with the observed distribution.
\label{fig_pdf_statis} 
}
\end{figure*}

\vspace{0.3em}
\noindent{\textbf{$\bullet$ The very-high-velocity cloud G165}}

The HI4PI survey  (refs.\citep{2009ApJS..181..398M,2011AN....332..637K,2016A&A...585A..41W,2016A&A...594A.116H})
provided so far the most detailed \HI~images of all-sky HVCs\citep{2018MNRAS.474..289W}.
VHVCs are dominated by the Magellanic Stream (MS) \citep{1972A&A....18..224W,1974ApJ...190..291M}
and the anticentre (AC) complex (refs.\citep{1985ApJS...58..143B,1988A&AS...75..191H,1991A&A...250..509W}). 
The VHVC component of the AC complex, denoted as G165 VHVC in this work, is
the most notable one in brightness and size on the sky (Extended Data Fig. 1), which stands out as an ideal object for investigating filament formation in extreme environments. 
Besides the VHVC component (-330 km s$^{-1}$ $<V_{\rm LSR}<$ -250 km s$^{-1}$), 
at the line of sight of G165 there is also an HVC component (-140 km s$^{-1}$ $<V_{\rm LSR}<$ -80 km s$^{-1}$) 
and a foreground component (-20 km s$^{-1}$ $<V_{\rm LSR}<$ 20 km s$^{-1}$). 
G165 HVC is part of the Cohen Stream, which was measured to have a distance bracket of 5.0 to 11.7 kpc\citep{2008ApJ...672..298W}.
In contrast, the H$\alpha$ brightness of G165 VHVC implies a distance ($d$) on the order of 15 kpc\citep{2008ApJ...672..298W}, the value we adopted.
The G165 VHVC is gravitationally unbound with a size of $d\times$100$\arcmin\sim 450$ pc, and a total gas mass of $\sim4\times10^5~M_{\odot}$ (see methods).
Limited by poor angular resolution ($\sim 16.2\arcmin$), the inner structures of the G165 VHVC were not resolved 
but did show clumpy features in the HI4PI survey.

\vspace{0.5em}
\phantomsection
\label{sec:results}

\vspace{0.3em}
\noindent{\textbf{$\bullet$ A WNM system with supersonic dynamics}}

We mapped the G165 VHVC using the Five-hundred-meter Aperture Spherical radio Telescope (FAST), 
with unprecedented angular resolution ($\sim 3\arcmin$, corresponding to a spatial resolution of 13 pc at a distance of $\sim$10 kpc), 
velocity resolution ($\sim 0.1$ km s$^{-1}$), and line sensitivity (60 mK per channel). Complex inner structures of the G165 VHVC are revealed in the integrated intensity map of the \HI~21 cm line (left panel in Fig. 1). 
Although different velocity subcomponents of the G165 VHVC are spectrally blended, they can be well 
separated through careful Gaussian decomposition (Fig. 1; 
see Sect. Methods for details).
These subcomponents are distributed in multiple layers in the position-position-velocity ({\it ppv}) cube (Fig. 2).
The velocity layers are interlaced with one another and exhibit large-scale velocity fluctuations (or velocity wiggles), similar to those seen in the molecular interstellar medium \citep{2020NatAs...4.1064H}.
The central velocities ($V_{\rm LSR}$) of the subcomponents can be divided into five groups ranging from $-325$ km s$^{-1}$ to $-240$ km s$^{-1}$ (Fig. 3a), though these velocity groups do not correspond one-to-one with the spatially separated gas layers.
The thickness of each velocity layer is estimated to be $\sim$100 pc (Sect. Methods).

The line widths (full width at half maximum; FWHM) of the velocity subcomponents are slightly 
larger than 20 km s$^{-1}$ (Fig. 3c), 
the value of the typical thermal line widths of the
WNM\citep{2003ApJ...586.1067H,2003ApJ...587..278W,2023ARA&A..61...19M}. 
Assuming a gas kinetic temperature of the G165 VHVC of 6000 K, the typical value of WNM\citep{2023ARA&A..61...19M},
the thermal dispersion is estimated as $\sim$7  km s$^{-1}$, 
yielding the 
transonic ($M_{\rm s}\sim 1$) nature of individual velocity subcomponents.
However, different velocity groups are separated by $\sim$20 km s$^{-1}$, suggesting the whole cloud could be in supersonic turbulence.
Considering that different gas layers are intersected, forming a dynamically correlated system (see also the next sections), all velocity groups can be treated as a whole system. The total dispersion of the centroid velocities ($V_{\rm LSR}$), a good tracer of the internal turbulent motions within a cloud, is up to $\sim$20 km s$^{-1}$, indicating a global $M_{\rm s}$ of $\sim$3. We note that the observed distribution of line widths (Fig. 3c) can be well reproduced by our MHD simulation of supersonic and trans-Alfv\'enic gas ($M_{\rm s}=3$ and $M_{\rm A}=1$). The same simulation also spans a large velocity range with several velocity groups (Fig. 3b), as the same in observations. 

The median values of column density ($7\times 10^{19}$ cm$^{-2}$) and  volume density (0.2 cm$^{-3}$) of the G165 VHVC (Sect. Methods) are
also similar to the typical values of the WNM\citep{2003ApJ...586.1067H}. 
The column density probability distribution function ($N$-PDF) of the decomposed subcomponents is higher-end suppressed (or lower-end enhanced), and
deviates from a lognormal shape (Fig. 3d).
The lognormal behavior of an $N$-PDF is usually explained by supersonic turbulence\citep{1997MNRAS.288..145P,2012ApJ...755L..19B,2015ApJ...811L..28B},
but the intermittency of a supersonic system caused by head-on collisions of strongly shocked gas
could lead to considerable deviations\citep{2010A&A...512A..81F,2017MNRAS.471.3753S}. The observed $N$-PDF can also be depicted by our MHD simulation models of supersonic 
and trans-Alfv\'enic gas ($M_{\rm s}=3$ and $M_{\rm A}=1$; see below).

To conclude, we consider the G165 VHVC as a WNM system with supersonic dynamics. This is significantly different from the majority of the WNM in Galactic plane or in HVCs (e.g., Smith cloud), where turbulence is subsonic to transonic \citep{XuSY_2019,2024AAS...24340224H,2007A&A...466..555H,2015ApJ...811L..28B}. \\


\noindent{\textbf{$\bullet$ Web-like filamentary structures}}

Filamentary structures in the G165 region, 
which cannot be resolved by HI4PI, are clearly seen by FAST on the \HI~moment 0 maps of its 
VHVC, HVC, and foreground components 
(Extended Data Fig. 2).
The HVC components show parallel filaments elongated along the Cohen Stream, similar to the famous Smith HVC \citep{2024AAS...24340224H}. In contrast, both the foreground and VHVC components show web-like filamentary structures.  
The foreground component has no clear boundary, 
while the VHVC shows clear boundaries defined by long and slim filaments.
Another difference between the foreground and VHVC components is that the filaments tend to be
in parallel mode on the channel maps of the foreground component, 
but stay intersected at different velocity slices of the VHVC components (Extended Data Fig. 2).
These comparisons indicate that the filamentary system of the G165 VHVC is really web-like rather than in parallel mode, which is substantially different from those in typical HVCs (e.g., Smith HVC \citep{2024AAS...24340224H}) and low/intermediate velocity \HI~clouds \citep{2014ApJ...789...82C,2023MNRAS.526.4345K}.

Filamentary structures in the G165 VHVC are visible in different layers of the velocity subcomponents (Fig. 2).
However, this may be misleading due to projection effects, which enhances the surface intensity at places 
tangent to the line of sight.
We developed a tool, {\it ClimbRidge}, based on a density ridge algorithm\citep{2015MNRAS.454.1140C,2016MNRAS.461.3896C}, to extract the
density ridges of G165 VHVC in both 2D and 3D cases (Sect. Methods).
{\it ClimbRidge} gives similar results as the well known tool {\it FilFinder} in 2D case (Extended Data Fig. 3),
and can be naturally applied to 3D ({\it ppv}) data to trace truly velocity-coherent filaments (Fig. 2 and
Extended Data Fig. 4).
Minimal Spanning Trees (MSTs) of the points on the density ridges were built (Extended Data Fig. 5; see Sect. Methods), and the
lengths of filament segments were measured (Fig. 3e).  As one can see, the distribution of segment lengths in our MHD simulation of supersonic and trans-Alfv\'enic gas ($M_{\rm s}=3$ and $M_{\rm A}=1$) remarkably resembles the observed one.

The filaments in the G165 VHVC are mainly distributed in three layers in {\it ppv} space (denoted as L1, L2 and L3 in Extended Data Fig. 4).
Each layer has an angular size larger than 2.5$\arcdeg$.
On the plane of (RA, DEC), the G165 VHVC can be divided into its
eastern, western, and southern parts. 
L1 is the main body of the western part.
L2 spans from the eastern part to the western part, with
a global velocity gradient of 0.2 km s$^{-1}$ per arcmin (0.05 km s$^{-1}$ pc$^{-1}$).
L3 extends from the western part to the southern part, with
a global velocity gradient of 0.3 km s$^{-1}$ per arcmin (0.07 km s$^{-1}$ pc$^{-1}$) (Extended Data Fig. 4).
The flare gas above L2, with velocity spreading to $>-240$ km s$^{-1}$, also has a filamentary structure. 
The filaments of the G165 VHVC are intersected with each other, building a filamentary network in the {\it ppv} space.

There are long and slim filaments that 
go through different layers, indicating that
different layers are indeed not independent and form a dynamically correlated system. 
Three representative filaments (denoted as B1, B2, and B3) are highlighted in the right panel of Fig. 2 and Extended Data Fig. 4. 
We examine these long filaments directly in the {\it ppv} cube (Extended Data Fig.~6), which reveals that their velocity structure is coherent, confirming that the density ridges trace the skeletons of the filamentary structures in {\it ppv} space.
Basic parameters of the three filaments are fitted and listed in Extended Data 
Table 1. 
The velocity gradient of the filaments is comparable with that of the gas layers ($\sim$0.1 km s$^{-1}$ pc$^{-1}$), contributing a velocity span of $>20$ km s$^{-1}$.
This also suggests that the filaments in G165 VHVC are supersonic.
The filaments of the G165 VHVC tend to be radially asymmetrical on the $l$-$r$ plane (for example see Extended Data Fig. 6).
The intensity difference between the two sides of the filament is comparable with the 
intensity enhancement of the central ridge (Extended Data Fig. 7). 
Our MHD simulations suggest that 
the asymmetrical density enhancement of filaments could be induced by supersonic turbulence (Fig. 3f).
Although the line width analysis and simulations (Fig. 3) suggest that the majority of the components of G165 VHVC are WNM, we cannot fully exclude the 
presence of CNM components, as unresolved thin layers of CNM may be induced by turbulent flows\citep{2005A&A...433....1A}.


Overall, we suggest that: (1) The G165 VHVC is a supersonic WNM system; (2) A filamentary 
network has already been established in such
an early-phase atomic system;
(3) The dynamic structure (several gas layers; multiple velocity groups; skewed $N$-PDF) and the filamentary network (including their asymmetrical radial profiles)
could be induced by supersonic shocks. The effect of self-gravity of the G165 VHVC onto itself and its filamentary substructures is negligible (Sect. Methods).
To conclude, filaments, and even web-like filamentary structures, can be well established in very low density but supersonic turbulent gas clouds (e.g., WNM in G165 VHVC), where self-gravity is not important.

\vspace{0.5em}
\phantomsection
\label{sec:discussion}


\noindent{\textbf{$\bullet$ Physical environment implication through numerical simulation}}

To further check the turbulent levels in the G165 VHVC, we conducted a series of isothermal MHD simulations with subsonic ($M_{\rm s}=0.5$), transonic ($M_{\rm s}=1$) and supersonic ($M_{\rm s}=3,~4$) turbulence (Sect. Methods). We do not have magnetic field information of the region 
but we expect that magnetic fields will have effects on the partially ionized WNM. Therefore, we also perform a parameter study with
$M_{\rm A}=$ 0.5, 1, and 2 as explained in the Methods section. 
Slim filaments with asymmetrical profiles
can be clearly produced by supersonic turbulence combined with trans/super-Alfv\'enic numbers (Extended Data Fig. 7). 
In {\it ppv} space, the simulated data shows a multi-layered pattern as in the observations (Supplementary Fig. 1). 
The velocity gradient and velocity wiggles seen in the 
observed gas layers (Fig. 2) can also be reproduced (Supplementary Fig. 1).
We argue that the
intermittency of supersonic turbulence could be responsible for the
asymmetrical profiles of filaments (Fig. 3f), which further contribute to the
lower-end enhancement of the $N-$PDF (Fig. 3d).

Gaussian decomposition and filament extracting procedures were applied to the simulated {\it ppv} cubes 
as to the observed one (Sect. Methods).
The three subsonic and transonic models ($M_{\rm s}=0.5, 1$) have a much narrower 
$N$-PDF than the supersonic models and the observed $N$-PDF (Extended Data Fig. 8).  
This indicates that we need a supersonically turbulent region to 
recreate the $N$-PDF of the observed region.  
The two slightly super-Alfv\'enic models ($M_{\rm A} = 2$) have longer low density tails than the observed $N$-PDF.  
The best fit to the observation is the model with $M_{\rm s} = 3$ and $M_{\rm A} = 1$.  
The best-fit simulation can also statistically reproduce the observed distribution of the line width, central velocity, and column density of the decomposed subcomponents (Figure 3a--d). 
It again supports that the G165 VHVC is a dynamically correlated supersonic and trans-Alfv\'enic system. We extracted filaments in the {\it ppv} space of the best-fit model and successfully
reproduced the observed distribution of the lengths of the filament segments (Figure 3e).
The filament network in G165 VHVC is most likely formed due to supersonic MHD turbulence.

The magnetic field strength of the G165 VHVC is then estimated to be
$\sim 4\ \mu G$ (see Sect. Methods). 
However, the lack of the low-density tail of the 
observed $N$-PDF could also be a result of detection sensitivity.  
Therefore, our simulations do not completely rule out the cases of somewhat weaker magnetic fields. We do not know how the G165 VHVC is bounded (or not). Besides a magnetic field, ram pressure and bulk motion may also 
contribute to anisotropic dynamics that
have not been considered by our simulation.  
At present, we can not exclude effects due to these anisotropic dynamics.
Instead,
our comparisons 
suggest that the supersonic MHD turbulence, 
in the sense of a zero-order approach,
can explain the multi-layer pattern and filament-network formation of G165 VHVC.

\vspace{0.3em}
\noindent{\textbf{$\bullet$ Origins of hierarchical filamentary structures}}

Hierarchical filamentary structures are ubiquitous in Galactic molecular clouds (MCs), manifesting as slim curves \citep{2013A&A...550A..38P}, hubs \citep{2013A&A...555A.112P}, and webs \citep{2010A&A...518L.102A}. In contrast, filamentary structures in atomic clouds tend to be simpler. However, in G165 VHVC, all these filamentary types -- slim curves, hubs, and webs -- are present, suggesting the possibility of forming hierarchical filamentary structures without the aid of gravity, as in MCs. The web-like filaments within G165 VHVC do not merely intersect due to projection effects. Instead, different filaments intersect in the {\it ppv} space, with mass assembling at the intersection knots (see filament B1, for example, in Extended Data Fig. 6). Different velocity layers are connected by slim filaments (Extended Data Fig. 4). Along the filaments of G165 VHVC, velocity gradients comparable to those in MCs (ranging from $<$0.1 km s$^{-1}$ pc$^{-1}$ to several km s$^{-1}$ pc$^{-1}$) have already been established. Note that velocity gradients along filaments in MCs are common \citep{Fernandez-Lopez_2014,2021ApJ...912..148L,2022MNRAS.514.6038Z}, and are often interpreted as gas flows driven by gravity. The results of this work suggest that, in addition to gravity, other mechanisms -- particularly MHD turbulence -- play an important role in the formation of hierarchical filamentary structures and in shaping the velocity gradients along filaments.

\clearpage
\phantomsection
\label{sec:methods}
\noindent{\large \textbf{Methods}}


\vspace{0.5em}
\noindent\textbf{1. FAST observation and data reduction}

The observations were conducted in on-the-fly (OTF) mode using the Five-hundred-meter Aperture Spherical radio Telescope 
(FAST)\citep{2011IJMPD..20..989N}. 
The illuminated aperture ($D$) of FAST is 300 meter for zenith angle $\theta_{ZA}<$26.4$\degr$,
and decreases by one third the value for $\theta_{ZA}=40\degr$\citep{2019SCPMA..6259502J}.
The tracking accuracy is 0.2$\arcmin$.
The $L$-band  (1--1.5 GHz) front end is a 19-beam dual-polarization (XX and YY) receiver (Supplementary Fig. 2).
It  has an FWHM beam size ($\Theta$)  of $\sim$3$\arcmin$ at 1.4 GHz,
a system temperature of less than 24 K \citep{2019SCPMA..6259502J},
and a beam efficiency for extended sources of 0.85 \citep{2019SCPMA..6259502J,2022A&A...658A.140L}.

The spectral backend provides 1024k (about one million) channels,
providing a velocity resolution of 0.1 km s$^{-1}$ at 1.4 GHz, much better than  that of the HI4PI survey ($\sim 1.5$ km s$^{-1}$)\citep{2016A&A...594A.116H}.
The analog-to-digital converter (ADC) of the FAST samples the data every 4 ns.
The time intervals related to the spectral backend are thus always integer multiples of 
$t_{\rm intrinsic}$=$2^{20}\times 4 \times  10^{-9}=0.004194304$ s.  
For our observations, the spectra were recorded every $dt_{\rm sample}=120 t_{\rm intrinstic} \sim 0.50331648$ s.
For calibration, the system periodically injects noise of  $T_{\rm cal}\sim 1.1$ K
every $4dt_{\rm sample}$  (with $\sim 1$ s of noise on and $\sim 1$ s of noise off).


G165 VHVC was covered by eleven OTF fields, each with an angular 
size of 1$^{\circ}$ $\times$ 1$^{\circ}$. 
The observation of each field is required to be completed within two hours to ensure a
proper range of  $\theta_{ZA}$. 
Thus, 
we adopt a maximum scanning speed of 30$\arcsec$ s$^{-1}$ and 
a scanning row separation ($l_{\rm sep}$) of 2.7$\arcmin$ (Supplementary Fig. 2).  
To achieve the criteria of Nyquist sampling,
the feed was rotated by 53.5$\degr$ clockwise.
The overhead time cost between two consecutive scanning rows is fixed to be 54 s.  
Each field requires an observing time of $\sim 1.1$ hours. In total, 13 hours were required
to cover all the eleven fields.

At present, there is no widely used data reduction pipeline for the FAST OTF observations. 
Below follows the description of our \href{https://gitee.com/liuxunchuan/fast_otf_datareduct}{data reduction procedure}.

\textbf{Gridding:} 
At present, FAST 
only records the start time ($t_{\rm start}$) of the  OTF observations. 
The sampling time ($t_{\rm sample}$) of the $n_{\rm th}$ spectrum can then be derived as $t_{\rm start}+ndt_{\rm sample}$. 
The accurate value (0.50331648 s) of $dt_{\rm sample}$  must be adopted.
If an approximate value of  0.5 s was adopted, 
the maximum deviation of the derived  $t_{\rm sample}$ would be 26 s,
leading to a maximum angular shift of $\sim 13\arcmin>4\Theta$. 
We then grid the data into $1\arcmin\times1\arcmin$ pixels according to the targeting position derived from $t_{\rm sample}$.   
The distribution of the effective integration time ($t_{\rm int}$) of the final cube is shown in 
Extended Data Fig. 1.


\textbf{Doppler correction:} 
For each spectrum, Doppler corrections for the rotation of Earth and the movement of Earth and Sun were applied 
according to its coordinate and $t_{\rm sample}$, using
the public Python package {\it PyAstronomy}.

\textbf{Calibration:} 
The spectra are dominated by very strong radio frequency interference (RFI) within the frequency range 1.15--1.3 GHz,
but are not significantly affected by  RFI around 1.4 GHz.
The spectra around 1.4 GHz were extracted for further reduction and analysis. 
For each data block of one minute observations, the average spectra of the data with the injected noise on and off were obtained and 
denoted as $y_{\rm on}$ and $y_{\rm off}$, respectively.
The spectral baselines ($y^{\rm base}$)  of $y_{\rm on}$ and $y_{\rm off}$ were  obtained through polynomial fitting on line-free channels,
and the difference between their spectral baselines is denoted as 
\begin{equation}dy^{\rm base} =y^{\rm base}_{\rm on}-y^{\rm base}_{\rm off}.\end{equation}
Each spectrum (with the $i_{th}$ one  denoted as $y_i$) in the data block was then calibrated
through
\begin{equation}
y_i^{\rm calied} = \frac{y_i}{dy^{\rm base}} T_{\rm cal}.
\end{equation}


\textbf{Split HI cube:} 
First-order polynomial fitting has been adopted to 
subtract the spectral baseline of the \HI~cube. The spectral baseline 
is mainly composed of the system temperature ($T_{\rm sys}$)
and the continuum emission   ($T_{\rm cont}$).
We applied 2D polynomial fitting on the point-source-free regions of the spectral baseline map
to further separate $T_{\rm sys}$ and $T_{\rm cont}$. The resultant $T_{\rm sys}$ is typically 19 K. The \HI~cube of G165 VHVC can be found in an online repository \citep{liu_2025_15347725}.

\textbf{Robust check:}
The continuum image is dominated by background quasars (Supplementary Fig.~3). We applied two-dimensional Gaussian fitting to point-like sources with peak intensities greater than 3~K, yielding full width at half maximum (FWHM) values typically in the range of 3--3.4~arcmin, comparable to the beam size. The point-like sources identified by FAST, which has a larger beam size, encompass those detected by the NRAO VLA Sky Survey (NVSS) \citep{1998AJ....115.1693C}. The continuum fluxes of the strong point-like sources measured by FAST and NVSS are consistent (Supplementary Fig.~3), with flux deviations typically smaller than \textbf{six} percent. We smoothed the \HI~cube from FAST (this work) and from HI4PI \citep{2016A&A...594A.116H} to the same resolution, and compared the spectra at a randomly selected location (Supplementary Fig.~4). The two spectra are nearly identical. We conclude that the \HI~cube we obtained is well calibrated and properly gridded.

\vspace{0.5em}
\noindent\textbf{2. Gaussian decomposition and filament extraction}

The integrated intensity (moment 0) maps  and channel maps of the FAST \HI~cube reveal clear filamentary structures 
for all three (VHVC, HVC, and foreground) components of G165 (Fig. 1 and Extended Data Fig. 2).
We apply Gaussian decomposition of the \HI~spectra of G165 VHVC using {\it scousepy} and develop the tool {\it ClimbRidge}
to extract the filamentary structures directly in the {\it ppv} cube. 

\textbf{Gaussian decomposition:}
We decomposed the velocity subcomponents of the G165 VHVC
using {\it scousepy}, a widely used multi-component spectral line decomposition tool\citep{Henshaw2016,Henshaw19}.  
The velocity range for fitting is from $-370$ to $-190$~km~s$^{-1}$.
At each pixel, G165 VHVC can be at most divided into three velocity subcomponents (Fig. 1).
This procedure decomposed the {\it ppv} cube into multiple layers of the fitted  (RA, DEC, $V_{\rm LSR}$) points (Fig. 2).
The residuals of the multiple Gaussian fitting are shown in Supplementary Fig. 5.

\textbf{Algorithm and code of ClimbRidge:} 
\href{https://gitee.com/liuxunchuan/sometools/tree/master/filament/ClimbRidge}{\it ClimbRidge} is a Python tool we have written
to extract the filaments in $n$-dimensional ($n$-d) space following a gradient-ascent method -- the density ridge algorithm\citep{2015MNRAS.454.1140C,2016MNRAS.461.3896C}.
This algorithm constructs a smooth surface (or ``potential'') by convolving a set of input data points with Gaussian kernels: \begin{equation} S(x_1,x_2,\dots,x_n) = \sum_j p_j \exp\left( -\sum_{i=1}^n \frac{(x_i - x_{i,p_j})^2}{\sigma_i^2} \right). \end{equation} Here, $p_j$ is the weight of the $j_{\rm th}$ input point, $x_{i,p_j}$ is its location along the $i_{\rm th}$ dimension, and $\sigma_i$ is the dispersion (corresponding to FWHM/$\sqrt{8\ln(2)}$) of the Gaussian kernel along the $i_{\rm th}$ axis. The algorithm then identifies the density ridges of the surface ($S$), defined as the locations where the eigenvector corresponding to the largest eigenvalue of the Hessian matrix of $S$ is aligned with the gradient of $S$. These density ridges are interpreted as filaments traced by the input data points. To numerically locate the density ridges, a set of test points is scattered throughout the $n$-dimensional space. Each test point ascends the surface by following the gradient direction projected onto the subspace spanned by the $n-1$ eigenvectors corresponding to the smaller eigenvalues of the Hessian matrix of $S$. We provide both CPU and GPU implementations of {\it ClimbRidge}.

\textbf{Test of ClimbRidge in 2D case:} In the 2D case, the Gaussian fitting results (RA, DEC, and integrated intensity) from {\it scousepy} are used as the input points for {\it ClimbRidge} to construct the 2D density ridges. To generate a smooth surface from the input data, a Gaussian kernel with a FWHM of 3.5$\arcmin$ in both the RA and DEC directions is applied, and the fitted integrated intensity is adopted as the weight of each input point.
The FWHM of the Gaussian kernel is chosen to be comparable to the beam size of FAST, in order to suppress noise that could cause the test points to become trapped in spurious local maxima, while still preserving sufficient spatial resolution.
For comparison, we also use {\it FilFinder} \citep{2015MNRAS.452.3435K} to detect filaments on the integrated intensity map (Extended Fig.~3). The filaments identified by both methods are similar, with {\it ClimbRidge} exhibiting greater sensitivity to weak structures. While {\it FilFinder} is limited to 2D analysis, {\it ClimbRidge} can be directly applied to data of arbitrary dimensionality.

\textbf{Extract filaments in 3D ppv cube:}  
In the 3D case, the Gaussian fitting results (RA, DEC, velocity, and integrated intensity) from {\it scousepy} are used as the input points for {\it ClimbRidge}. 
The FWHM of the Gaussian kernel is adopted to be 3.5$\arcmin$ in both the RA and DEC directions, and 15~km~s$^{-1}$ in the velocity direction. Here, a FWHM of 15~km~s$^{-1}$, slightly smaller than the typical line width, is adopted to ensure that the scattered points climb toward different velocity layers. A total of 15,000 test points are scattered throughout the {\it ppv} space and ascend along the projected gradient. The resulting density ridges are shown in Fig.~1 and Extended Data Fig.~4.

\textbf{MST analysis:} 
We have developed a tool, \href{https://gitee.com/liuxunchuan/sometools/tree/master/filament/Fil3d}{\it fil3d}, 
to analyse the ridge points produced by {\it ClimbRidge}, including (1)
trimming out the isolated points, (2) minimal spanning tree analysis,
(3) and find out the shortest path between two given  points. 
We have built the minimal spanning trees (MSTs) 
to connect the density-ridge points  (Extended Data Fig. 5), 
using the public Python package \href{https://pypi.org/project/mistree/}{\it MiSTree},
for both the 2D and 3D cases.
To build the 3D tree, the
Euclidean distance is normalized by the units of (pixel, pixel, km s$^{-1}$)
in the three directions.
A segment is defined as the path that connects two neighbouring nodes.
A branch is defined as one or more connected segments.
We also redistributed the density-ridge points through randomly 
rearranging their RA, DEC and $V_{\rm LSR}$, and then built a random tree. 
The maximuum segment length of the random tree is
obviously shorter than that of the 3D tree (Fig. 3).
The 2D and 3D trees of G165 VHVC are qualitatively similar to each other (Extended Data Fig. 5).
The lower end of the segment-length PDF of the 3D tree
is more flattened than that of the 2D tree. It implies that the density ridges in the {\it ppv} cube can trace shorter filamentary branches, especially in crowded regions such as branch B3 (Extended Data Fig.~4). This may be because, in the 3D case, dynamically unrelated branches that appear connected in projection can be more effectively disentangled.


\vspace{0.5em}
\noindent\textbf{3. Derive physical parameters}

There are three main velocity layers, with flare gas extending to velocities greater than $-$240 km s$^{-1}$ (Extended Data Fig.~4). Five velocity groups are fitted from the central velocity distribution (Fig.~3). 
Flare gas contributes to a number of velocity groups in addition to the main velocity layers.
Therefore, we adopt an effective layer number of 4 (the mean of the numbers of main velocity layers and velocity groups) to estimate the layer thickness. Adopting a total thickness of the G165 VHVC of 1.5$\arcdeg$ (a value comparable to its angular size), the layer thickness ($H_{\rm layer}$) is roughly estimated to be
\begin{equation}
H_{\rm layer} = 1.5\arcdeg \times \frac{d}{4} \sim 100\ {\rm pc},
\end{equation}
where $d$ is the distance between G165 VHVC and us.
The \HI~emission of the G165 VHVC is weak and optically thin (Fig. 1). 
The \HI~column density can be directly calculated through \citep{2003ApJ...585..823L,2023ARA&A..61...19M}
\begin{equation}
N({\mbox{\HI}}) = 1.9 \times 10^{18} \frac{T_{\rm ave}\Delta V}{\rm K\ km\ s^{-1}} \ {\rm cm}^{-2},
\end{equation}   
where $\Delta V$ is the FWHM velocity width and $T_{\rm ave}$ is the averaged peak brightness temperature. Note that we have accounted for the conversion factor of 1.064 from \( T_{\rm ave} \Delta V \) to the integrated intensity of a Gaussian profile.
The median value of the \HI~integrated intensity ($T_{\rm ave}\Delta V$) is 37 K km s$^{-1}$, yielding a median column density of
$7\times 10^{19}$ cm$^{-2}$. It corresponds to a typical global volume density, $n_{\rm H}$, of 0.2 cm$^{-2}$. 
The linear mass ($\mathcal{L}$) of the \HI~filament can then be derived through
\begin{equation}
\mathcal{L} = 6.6\frac{d}{\rm 15\ kpc}\frac{\Delta r}{5\arcmin}\frac{\Delta V}{\rm 20\ km\ s^{-1}}\frac{T_{\rm ave}}{\rm K}\ {\rm M}_{\sun}\ {\rm pc}^{-1}, \label{eq_lm}
\end{equation}
where $\Delta r$ is the FWHM angular width of the filament.

The gravity-related critical density of a filament can be estimated through \citep{2021ApJ...912..148L} 
\begin{equation} 
\mathcal{L}_c = \frac{2\sigma_v^2}{G},
\end{equation}
where $G$ is the gravitational constant, and $\sigma_v$ is the 
velocity dispersion (FWHM/$\sqrt{8\ln(2)}$). 
Inserting the typical values of Extended Data Table 1 into Eq. \ref{eq_lm}, we get $\mathcal{L}<10\ M_{\sun}$ pc$^{-1}$.
Taking $\sigma_v$ as 10 km s$^{-1}$, we get $\mathcal{L}_c\sim 46\,000$ $M_{\sun}$ pc$^{-1}$, much larger than the value of $\mathcal{L}$. 
The total \HI~flux of the G165 VHVC is 350 K km s $^{-1}$ deg$^2$, 
corresponding to a total mass of 
\begin{equation}
M = m_{\rm H}d^2\int N({\mbox{\HI}}) {\rm d}\Omega \sim 4\times10^5 \left(\frac{d}{\rm 15\ kpc}\right)^2\ M_{\sun},
\end{equation}
where $m_{\rm H}$ is the mass of hydrogen atom and $\Omega$ is the angular coverage. 
The virial mass of the G165 VHVC is estimated to be
\begin{equation}
M_{\rm vir} \sim \frac{R\sigma_v^2}{G}\sim 1\times 10^7\ M_{\sun}.
\end{equation}
Here, $R$ is the physical size of G165 VHVC,  estimated to be $R=d\times 1.5\degr$.
$M_{\rm vir}$ is also much larger than $M$, implying that the self-gravity of G165 VHVC can be ignored. 


\vspace{0.5em}
\noindent\textbf{4. Ideal MHD simulation}

We use the multi-physics code ORION2 that is based on the Godunov scheme \citep{2012ApJ...745..139L,2021JOSS....6.3771L} to simulate the observed G165 VHVC.  
The ideal MHD module in ORION2  is based on the dimensionally unsplit corner transport 
upwind scheme\citep{Colella1990} incorporating the constrained transport framework\citep{1988ApJ...332..659E,2008Natur.454...71S}.  
ORION2 also has the adaptive refinement capability but we shall only use a single $512^3$ grid 
in our simulations here.  From observation, the G165 VHVC is likely supersonic with a $M_{\rm s}$ of $\sim$3.
Besides this value,
we have also performed a few more models at  $M_{\rm s}$ of 0.5 (subsonic), 1.0 (transonic), and 4.0 (even more supersonic) for comparison. 
We do not have magnetic field information of the region but we expect a magnetic field will affect the ionized WNM.  
Therefore, we perform a parameter study of $M_{\rm A}$ with values 0.5, 1, and 2 using turbulence boxes with periodic boundary conditions. 
Here, $M_{\rm A}=\sigma_{\rm turbulence}/V_{\rm A}$ with the Alfv\'en speed $V_{\rm A}=2.18/\sqrt{n_{\rm H}/{\rm cm}^{-3}}\times B/\mu {\rm G}$ km s$^{-1}$ for an \HI~system.
We assume that the VHVC is exposed to a uniform ultraviolet field and has a constant temperature.  
The overall volume density of G165 VHVC is smaller than the typical phase-transition density of WNM \citep{2003ApJ...587..278W}. The enhanced cooling rate in the most central regions of the filamentary hubs may lead to the existence of a colder HI phase, but it should not significantly influence the global dynamics. 
Therefore, we use the isothermal equation of state in the simulation.  
As discussed above,  
we also ignore gravity in the simulations.  
The simulated models are thus scaleless
and only controlled by $M_{\rm s}$ and $M_{\rm A}$.
In each model, we start with an initially uniform density and magnetic field along the z-axis.  
We use the recipe of MacLow\citep{1999ApJ...524..169M} in driving the turbulence boxes with a pure solenoidal velocity 
field continuously at the largest scale (k=1-2) for two crossing times $t_f = L/v_{\rm rms}$, 
where $L$ is the length of the box and $v_{\rm rms}$ is the root-mean-square turbulence velocity.  
For $M_{\rm s} = 1$, $v_{\rm rms}$ is the sound speed $c_s$. For comparison with observation, 
we only use the simulation data at $2t_{f}$.

To mimic a random observational  direction, we inspected the simulated models at the viewing angle of $30^{\circ}$ from the initial magnetic field orientation (z-axis), and produced the simulated {\it ppv} cubes under the optically thin limit (Supplementary Fig. 1). Further, we smoothed the simulated {\it ppv} cubes 
to have a map size (256 pixel) similar to that of observation. 
Gaussian decomposition was applied to the simulated {\it ppv} cubes, as  to the observed data.
For the best-fit model ($M_{\rm s}=3$ and $M_{\rm A}=1$), we further extracted its filamentary structures in the {\it ppv} space using {\it ClimbRidge} (Extended Data Fig. 7). 

\clearpage
\noindent\textbf{Data Availability}
The FAST \HI~cube of G165 VHVC, smoothed to a velocity resolution of $\sim$1~km~s$^{-1}$, can be found at \url{https://zenodo.org/records/15347725}.
Owing to data size, the full data used for this study are available from the corresponding authors upon reasonable request. 


\vspace{0.2em}
\noindent\textbf{Code Availability}

The FAST OTF data reduction script we developed for this work can found at \url{https://gitee.com/liuxunchuan/fast_otf_datareduct}.
The filament extracting code we developed for this work, {\it ClimbRidge} and {\it fil3d}, can be found at 
\url{https://gitee.com/liuxunchuan/sometools/tree/master/filament/}.

The MST analysis tool is available at \url{https://pypi.org/project/mistree/}.
The compared filament extracting method {\it FilFinder} can be found at \url{https://github.com/e-koch/FilFinder}. 
Image processing package {\it scikit-image} can be found at \url{https://pypi.org/project/scikit-image/}.
The ORION2 code is available at \url{https://joss.theoj.org/papers/10.21105/joss.03771}.

\vspace{0.2em}
\noindent\textbf{Acknowledgements}
We wish to thank the staff of the FAST for their kind
help provided during the observations.
X.L. and T.L. have been supported by the National Key R\&D Program of China (No. 2022YFA1603100).
X.L. has also been supported by
the Strategic Priority Research Program of the Chinese Academy of Sciences  under Grant No. XDB0800303.
T.L. acknowledges National Natural Science Foundation of China (NSFC) No. 12017061 and 12122307.
T.L. also acknowledges the supports  by the Tianchi Talent Program of Xinjiang Uygur Autonomous Region and the PIFI program of Chinese Academy of Sciences through grant No. 2025PG0009.
P.L.  acknowledges NSFC No. 1241101426.
S.Q. acknowledges NSFC No.12033005.
N.T. acknowledges the supports by the NSFC No. 12473023, and by the University Annual Scientific Research Plan of Anhui Province (No. 2023AH030052 and 2022AH010013).
This research was carried out in part at the Jet Propulsion Laboratory, which is operated by the California Institute of Technology under a contract with the National Aeronautics and Space Administration (80NM0018D0004).
We thank the anonymous reviewers for providing feedback that helps to solidify this work.

\vspace{0.2em}
\noindent\textbf{Author contributions}

X.C. L. was the principal investigator of the FAST project. He led the data reduction, filament-extracting code development, scientific analysis and wrote the manuscript.
T. L. contributed to the FAST proposal, the interpretation of results and writing of the manuscript. P.S. L. conducted the MHD simulation. All authors participated in the discussion of results,  and revision of the manuscript.


\vspace{0.2em}
\noindent\textbf{Competing interests}

The authors declare no competing interests.


\clearpage
\phantomsection
\noindent{\large \textbf{Extended Data figures}}

\setcounter{figure}{0}
\setcounter{table}{0}
\renewcommand{\figurename}{Extended Data Fig.}
\renewcommand{\figureautorefname}{Extended Data Fig.}
\renewcommand{\tablename}{Extended Data Table}
\renewcommand{\tableautorefname}{Extended Data Table}


\begin{figure*}[!h]
\centering
\hspace{0.006\linewidth}
\includegraphics[width=0.95\linewidth]{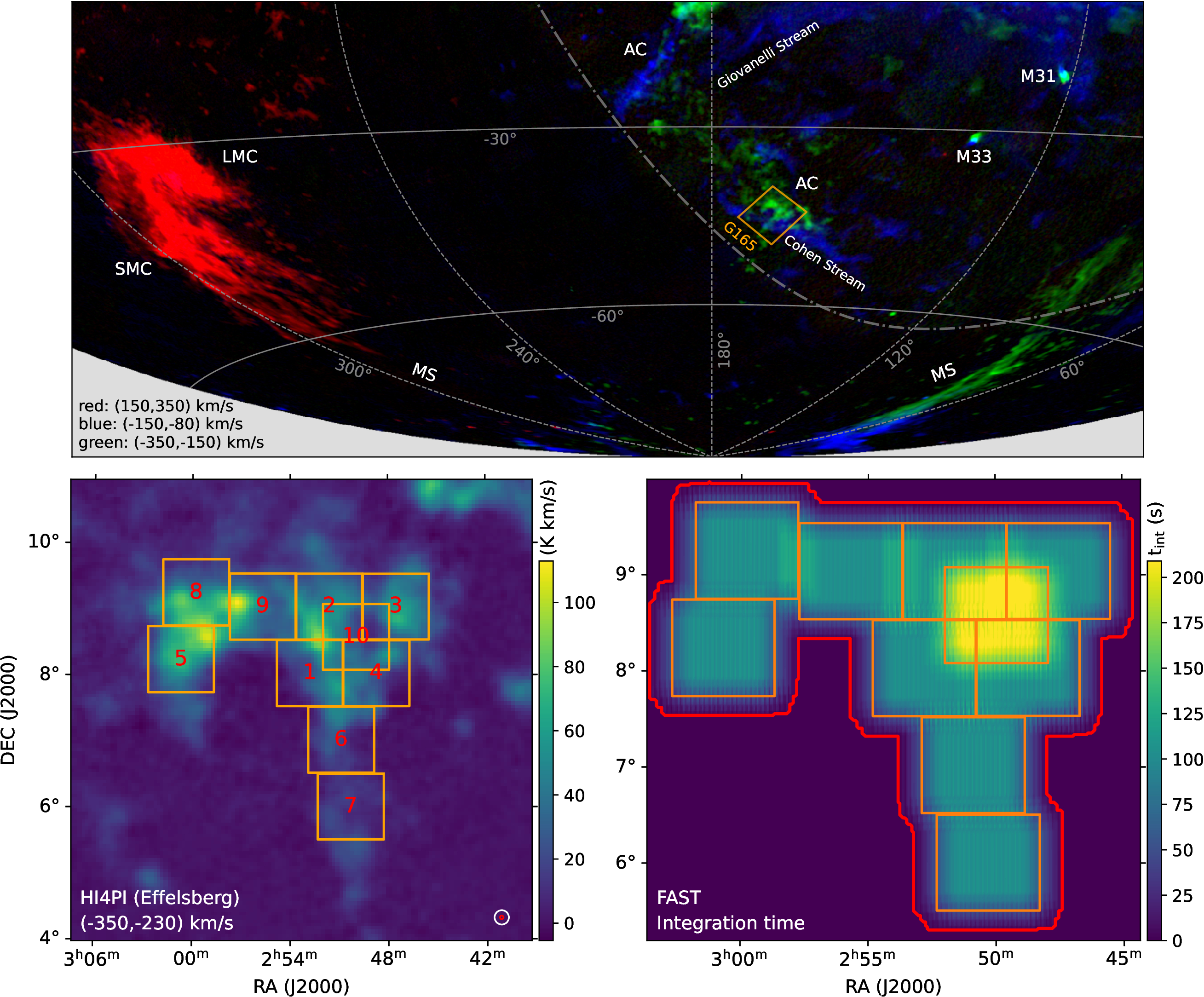}
\caption{
Location and scanning pattern of the G165 VHVC.
Upper panel: Color map composed by the H{\sc i} emission of three different
velocity ranges from HI4PI survey $^{41}$. 
Prominent HVC complexes (AC complex, Cohen Stream, Giovanelli Stream, and MS) and  external galaxies
(LMC, SMC, M33, and M31) are labeled.
Lower left panel: Zoom-in image of G165 VHVC from the HI4PI survey.
The orange rectangles ($1\degr\times 1\degr$ each in size) 
mark  the ten on-the-fly observation fields of FAST. 
The white and red circles represent the beam size of HI4PI and FAST, respectively. 
Lower right panel: Effective integration time of the FAST observations.
\label{fig_sampleoverview}
}
\end{figure*}

\begin{figure*}[!h]
\centering
\includegraphics[width=1.0\linewidth]{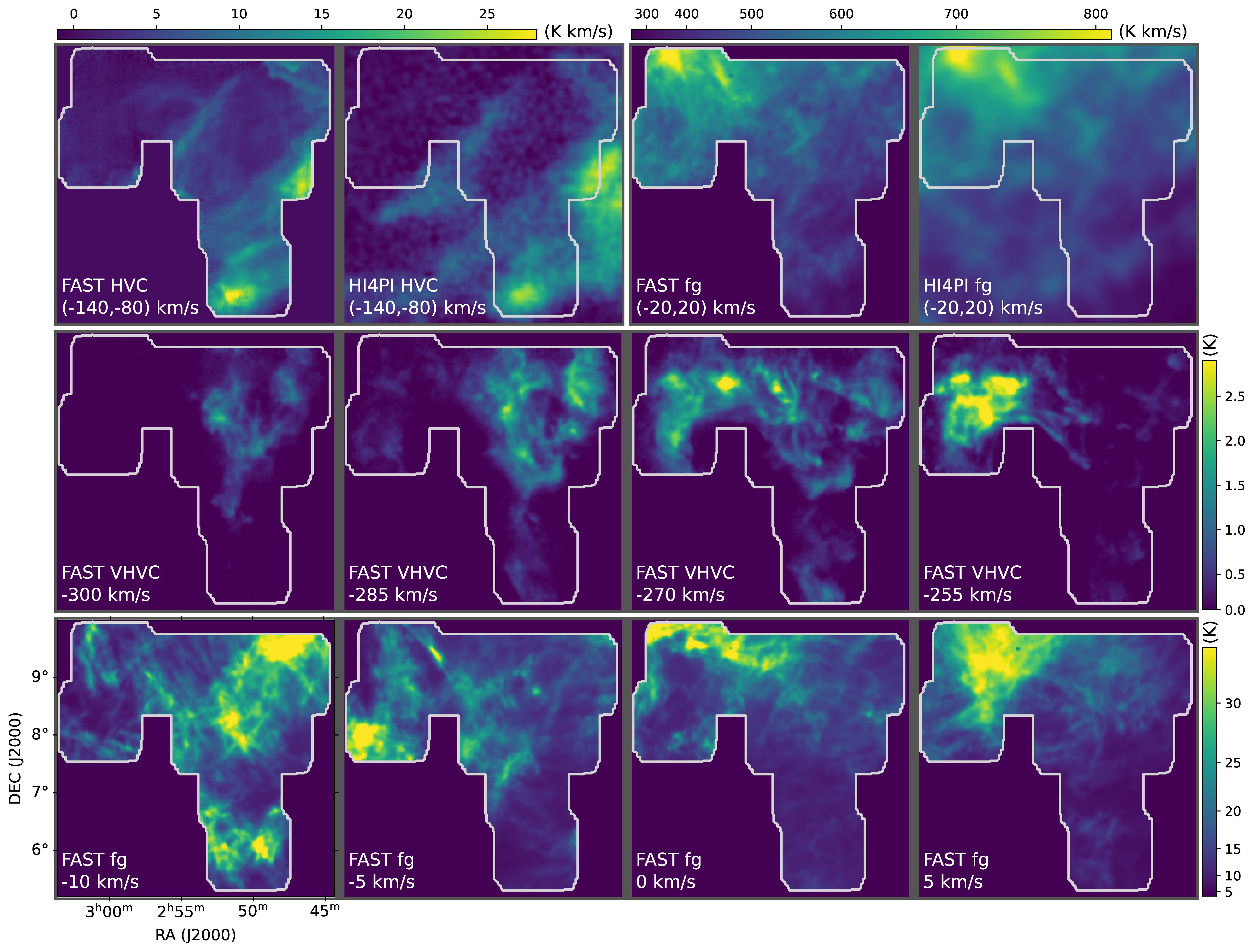}
\caption{
Comparison between the FAST observations and archival HI4PI data.
\textbf{Upper panels}: The integrated intensity maps of the HVC and foreground (fg) components of G165 by FAST and HI4PI.
See Fig. 1 and Extended Data Fig. 1 for the integrated intensity maps of the VHVC component.
\textbf{Center panels}: Channel maps of G165 VHVC by FAST with velocity resolution smoothed to 2 km s$^{-1}$ (about one fiftieth of the velocity span).
\textbf{Lower panels}: Channel maps of the foreground of G165 by FAST with velocity resolution smoothed to 0.3 km s$^{-1}$ (about 
one fiftieth of the velocity span).
\label{fig_HI_mom0}}
\end{figure*}

\begin{figure*}[!h]
\centering
\includegraphics[width=0.95\linewidth]{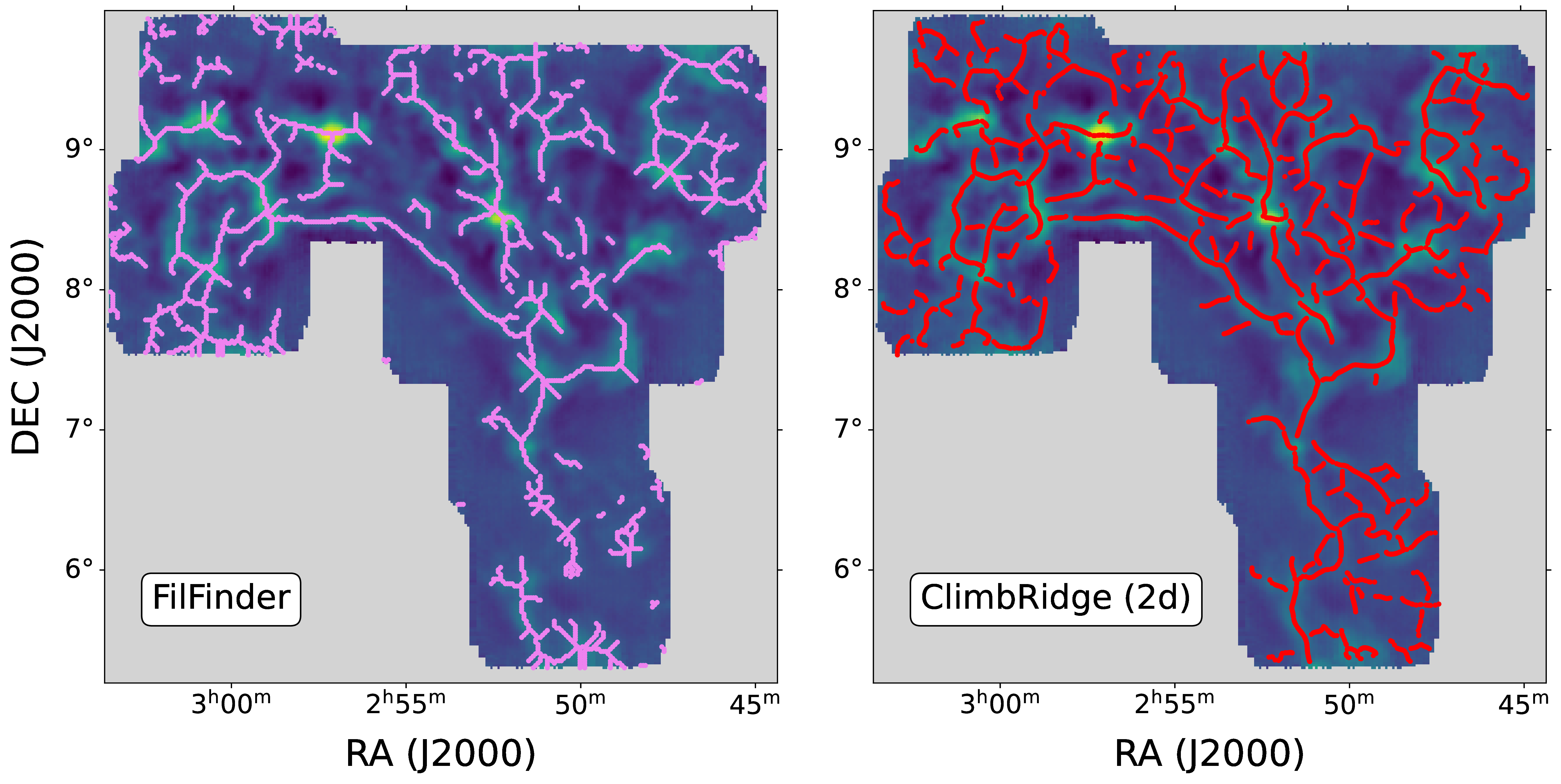}
\caption{Comparison between the filaments of the G165 VHVC extracted using different algorithms. The left panel shows the result from {\it FilFinder}, and the right panel shows the result from {\it ClimbRidge}. The background in both panels is the integrated intensity map of the G165 VHVC, processed with the {\it Sharpen} operation from the {\it scikit-image} Python package to enhance the filamentary structures (see Sect.~Methods). \label{fig_filfinder_CR}}
\end{figure*}

\begin{figure*}[!h]
\centering
\includegraphics[width=0.95\linewidth]{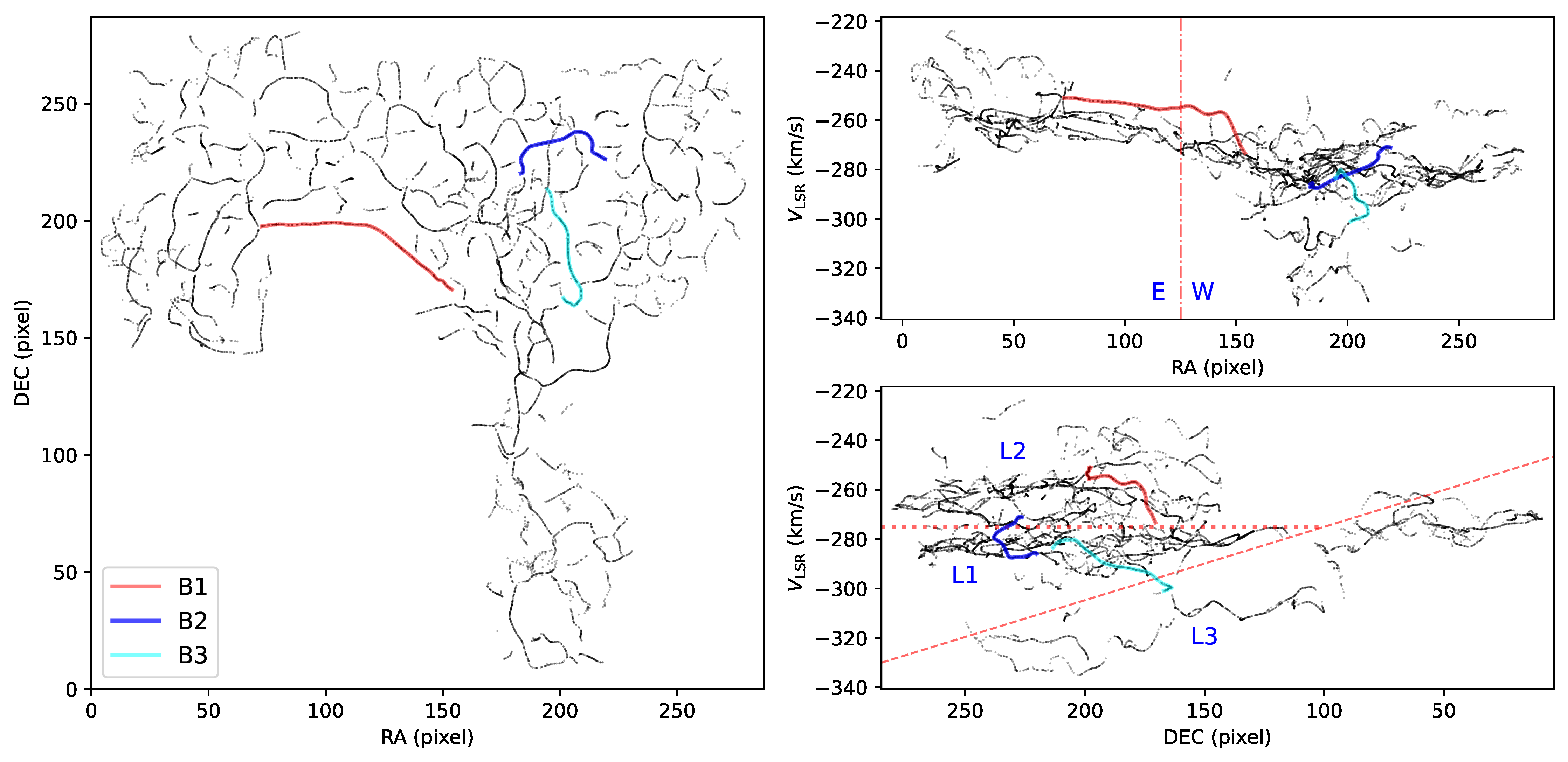}
\caption{
Filaments extracted using {\it ClimbRidge} in the {\it ppv} cube of the G165 VHVC, viewed from different projection angles. 
The projections are along the three axes: RA, DEC, and $V_{\rm LSR}$. 
See Fig.~2 for off-axis projections. 
Three representative filaments, B1, B2, and B3, are shown in different colors (see Fig.~2). 
In the bottom-right panel, the red dotted and dashed lines separate the three velocity layers (L1, L2, and L3) of the G165 VHVC. 
The red dashed line has a slope of 0.3\,km\,s$^{-1}$ per arcmin.
\label{fig_ridge_3angles}
}
\end{figure*}

\begin{figure*}[!h]
\centering
\includegraphics[width=0.99\linewidth]{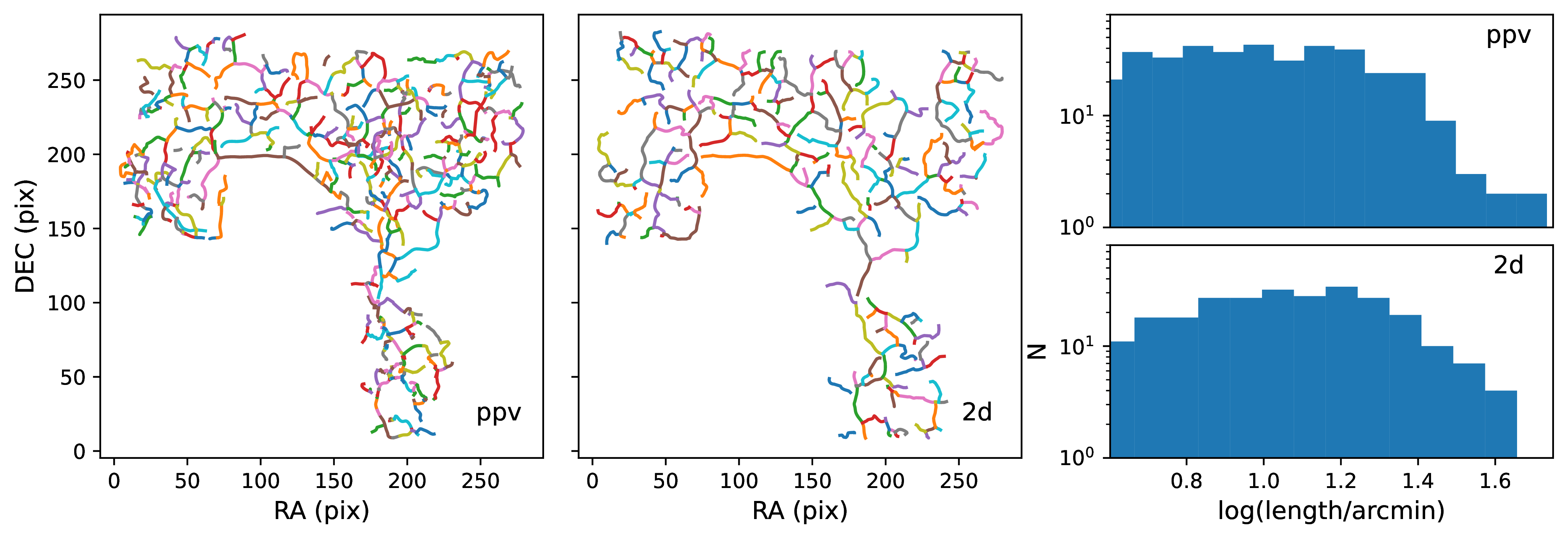}
\caption{
Minimal spanning tree (MST) analysis of the H{\sc i} segments of the G165 VHVC. 
\textbf{Left and Middle panels}: The MSTs of the density ridge points extracted by {\it ClimbRidge} in the 3D {\it ppv} cube (left) and on the 2D integrated intensity map of the G165 VHVC. 
Different colors represent different segments connecting every two neighboring nodes of the MSTs. 
\textbf{Right panel}: The PDFs of the segment lengths. 
\label{fig_mst}
}
\end{figure*}

\begin{figure*}[!h]
\centering
\includegraphics[width=0.98\linewidth]{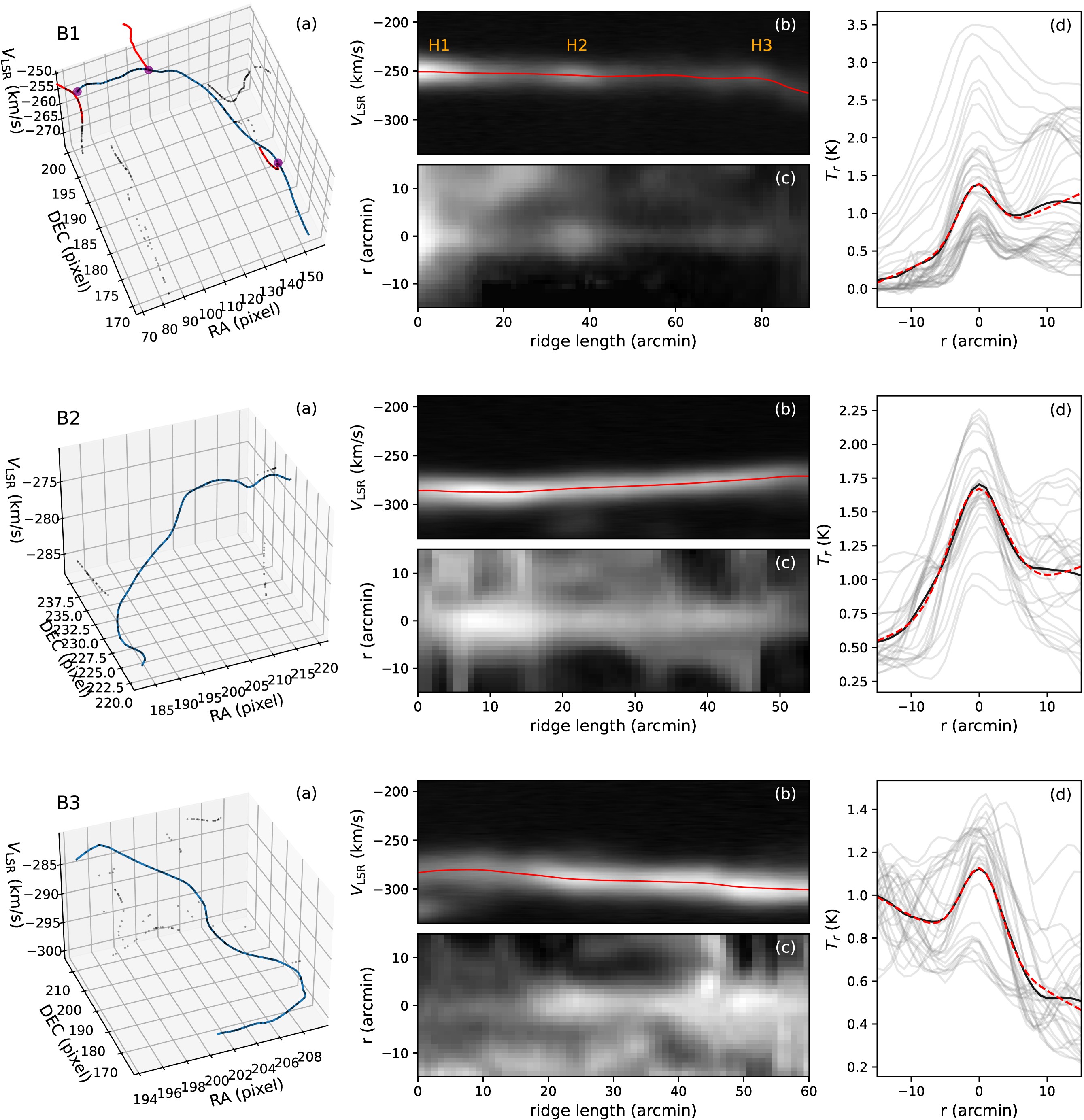}
\caption{
Three example H{\sc i} filaments of the G165 VHVC. 
(\textbf{a}) Zoom-in images of the three filaments shown in Extended Data Fig. 4. 
(\textbf{b}) Slice in the $l$--$V_{\rm LSR}$ plane, where $l$ is the length along the ridge. 
(\textbf{c}) Slice in the $l$--$r$ plane, where $r$ is the radial offset from the ridge. 
(\textbf{d}) Radial intensity profiles (gray lines) and the mean profile (black line). 
The red dashed line shows the fitting result of the black line using a Gaussian function plus a first-order polynomial.
\label{fig_3branches}
}
\end{figure*}

\begin{figure*}[!h]
\centering
\includegraphics[width=0.98\linewidth]{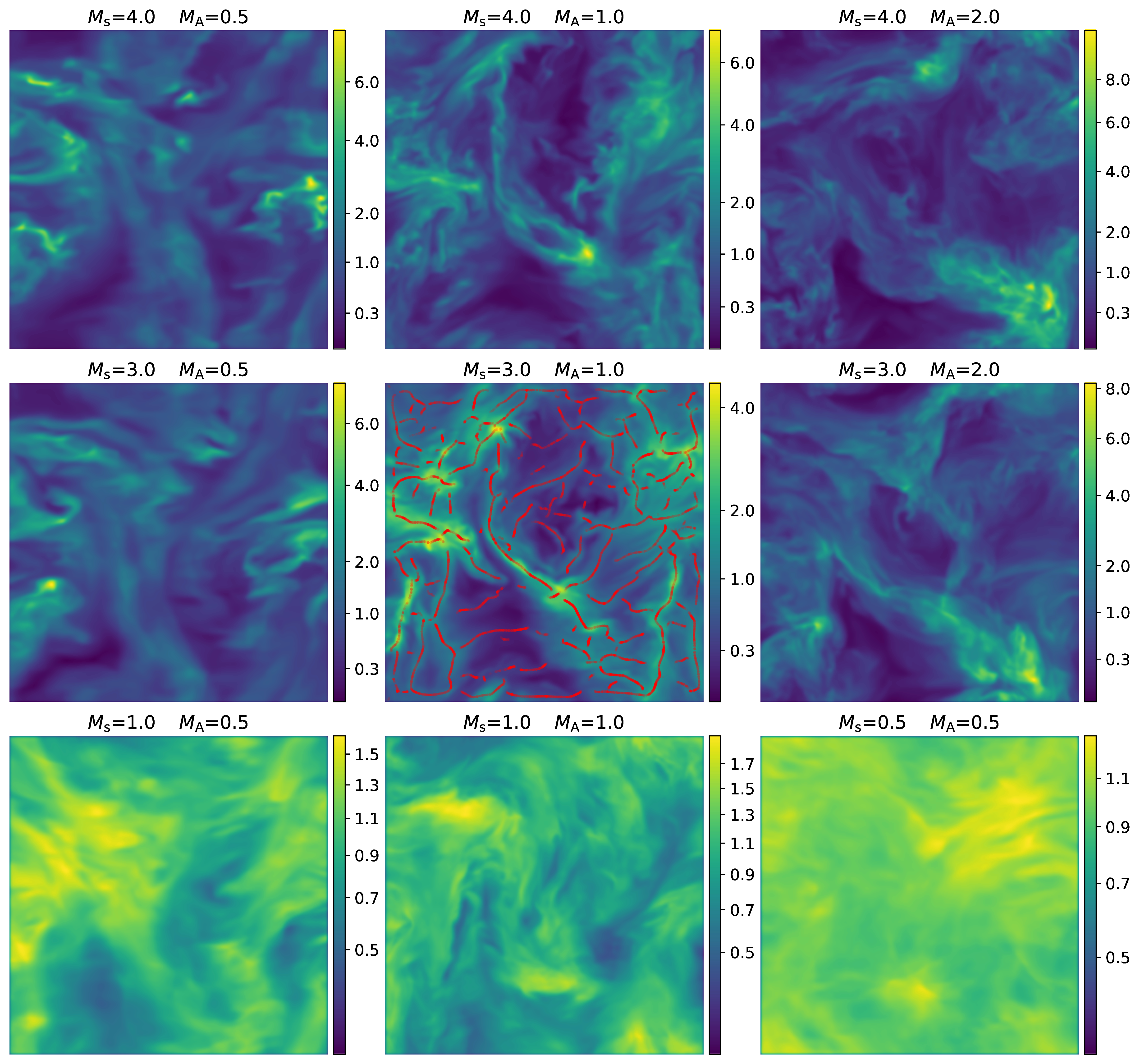}
\caption{The integrated intensity maps of the MHD simulation. In each panel, 
the integrated intensity has been normalized by its mean value.  The filaments of the best-fit model ($M_{\rm s}=3$ and $M_{\rm A}=1$) 
were extracted in the {\it ppv} cube (Supplementary Fig. 1) using {\it ClimbRidge} and shown as red lines. \label{fig_sim} }
\end{figure*}

\begin{figure*}[!h]
\centering
\includegraphics[width=0.9\linewidth]{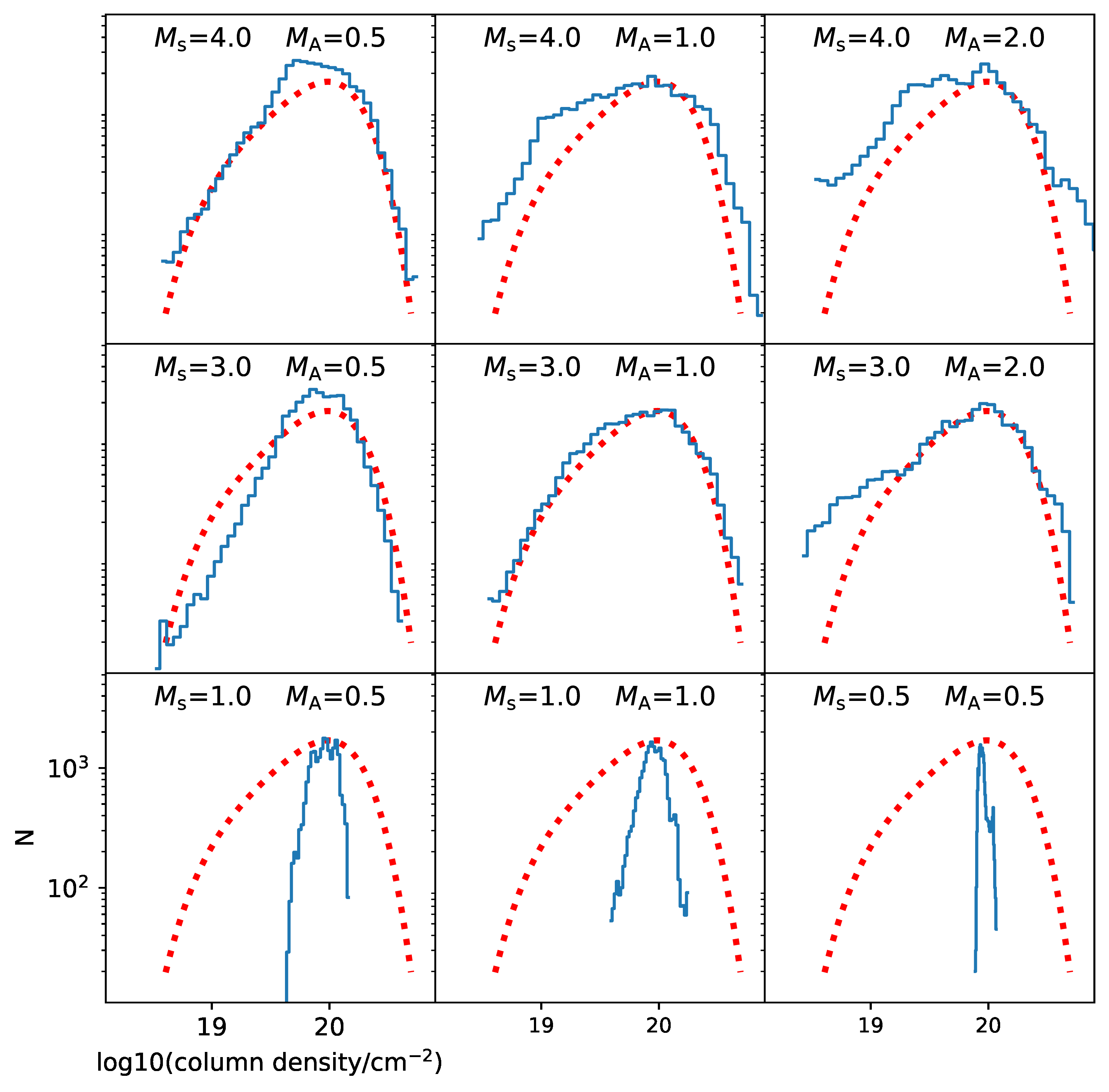}
\caption{The column density PDFs of the decomposed Gaussian components of the simulated {\it ppv} cubes.
The blue lines are the simulated PDFs. The red dotted line represents the fourth-order polynomial fitting of the observational result for the G165 VHVC (Fig. 3). \label{fig_cmp_pdf} }
\end{figure*}

\clearpage
\phantomsection
\noindent{\large \textbf{Extended Data tables}}

\begin{table}[h]
\caption{Fitted parameters of the typical filaments.\label{tab_ridge_4proj}}
\centering
\includegraphics[width=0.8\linewidth]{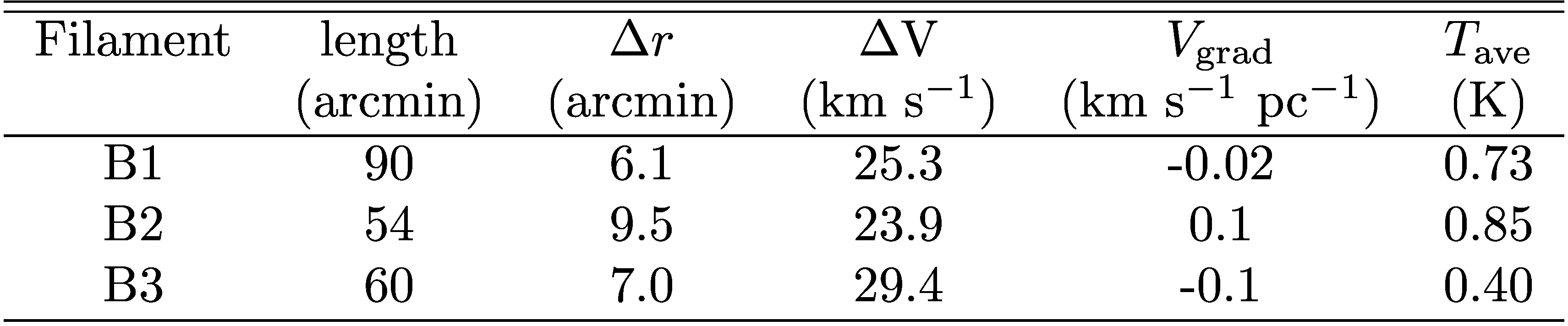}\\
{\small 
Note: The listed parameters include the angular length, velocity gradient ($V_{\rm grad}$), FWHM angular width ($\Delta r$), 
FWHM velocity width ($\Delta V$), and the averaged peak brightness temperature ($T_{\rm ave}$) of the typical filaments (Extended Data Fig. 6).
Adopting a distance of 15 kpc, the converting factor between the linear scale (pc) and angular scale (arcmin) is 4.36 pc arcmin$^{-1}$. 
}
\end{table}

\clearpage
\phantomsection
\noindent{\large \textbf{Supplementary figures}}

\setcounter{figure}{0}
\renewcommand{\figurename}{Supplementary Fig.}
\renewcommand{\figureautorefname}{Supplementary Fig.}

\begin{figure*}[h]
\centering
\includegraphics[width=0.99\linewidth]{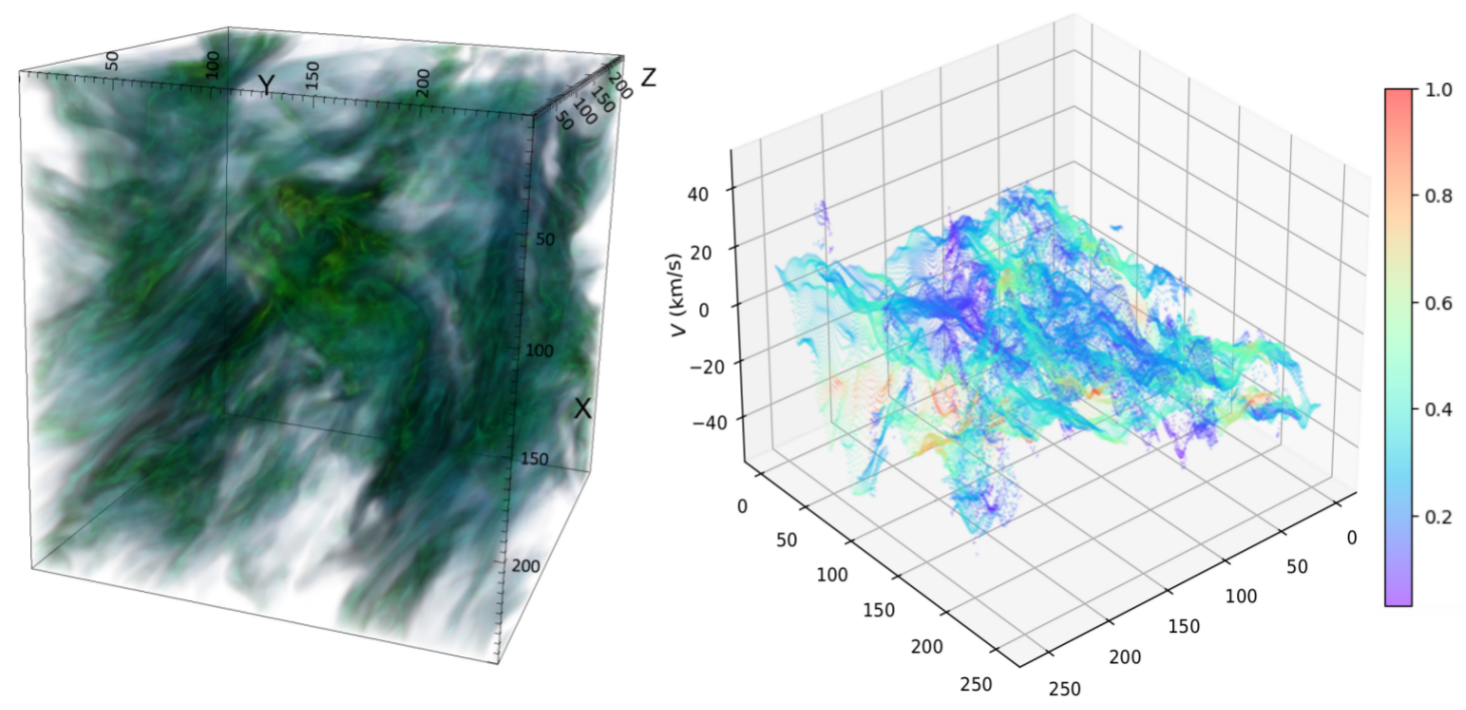}
\caption{
The 3D images of the best-fit simulated model.
\textbf{Left}: The volume-rendering cube (left panel) of the best-fit model (Extended Data Fig.~7).
\textbf{Right}: The distribution of the decomposed Gaussian components in the {\it ppv} cube of simulation. The color represents the integrated intensity. Note that the simulation is scale-free and only the relative intensity is meaningful. 
\label{simu_ppv_fig}
}
\end{figure*}

\begin{figure*}[!h]
\centering
\includegraphics[width=0.99\linewidth]{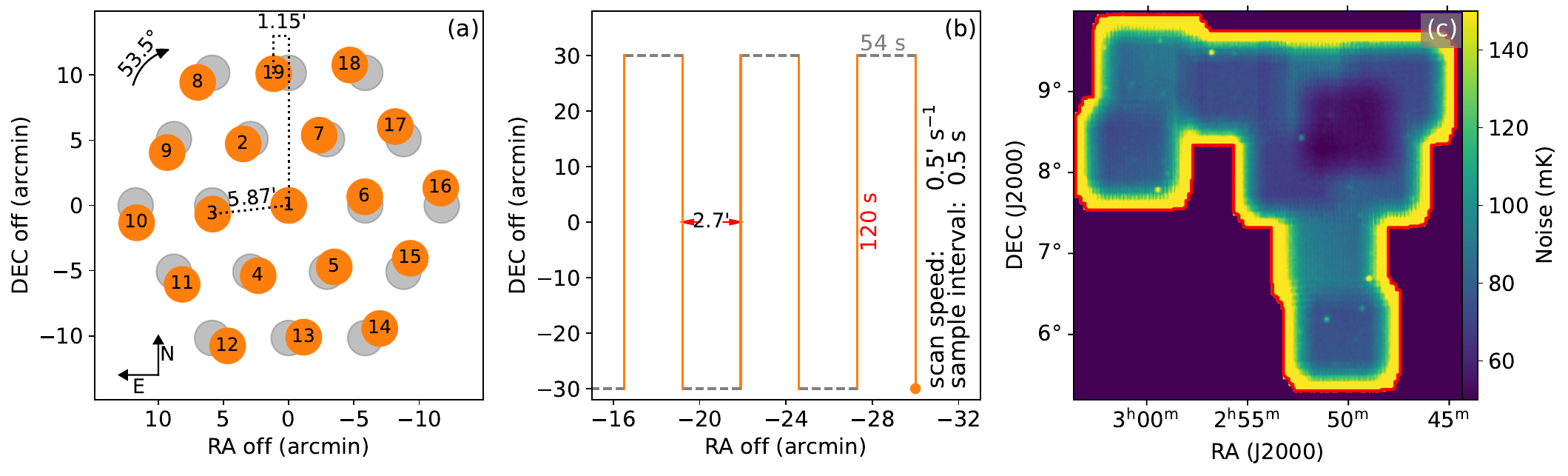}
\caption{
Detailed scanning pattern of G165 VHVC using the FAST. 
(\textbf{a}): The relative locations of the 19 beams of the $L$-band receiver of the FAST. 
For a position angle of the receiver ($\theta_{r}$) of 0$\degr$, the beams are distributed as gray circles, with beam 2 located east of beam 1. 
The beams rotate clockwise as $\theta_{r}$ increases. Our observations adopted a $\theta_{r}$ of 53.5$\degr$ (orange circles). 
(\textbf{b}): The scan pattern (orange lines) of our on-the-fly (OTF) observations. An overhead time of 54 seconds was consumed between two consecutive scanning rows. 
The orange dot marks the starting point at the location (-30$\arcmin$, -30$\arcmin$) relative to the field center. 
(\textbf{c}): The noise map of the \HI~cube of G165 VHVC (at a spectral resolution of 0.1 km\,s$^{-1}$) enclosed by the red contour. 
\label{fig_otfoverview}
}
\end{figure*}

\begin{figure*}[!h]
\centering
\includegraphics[width=0.99\linewidth]{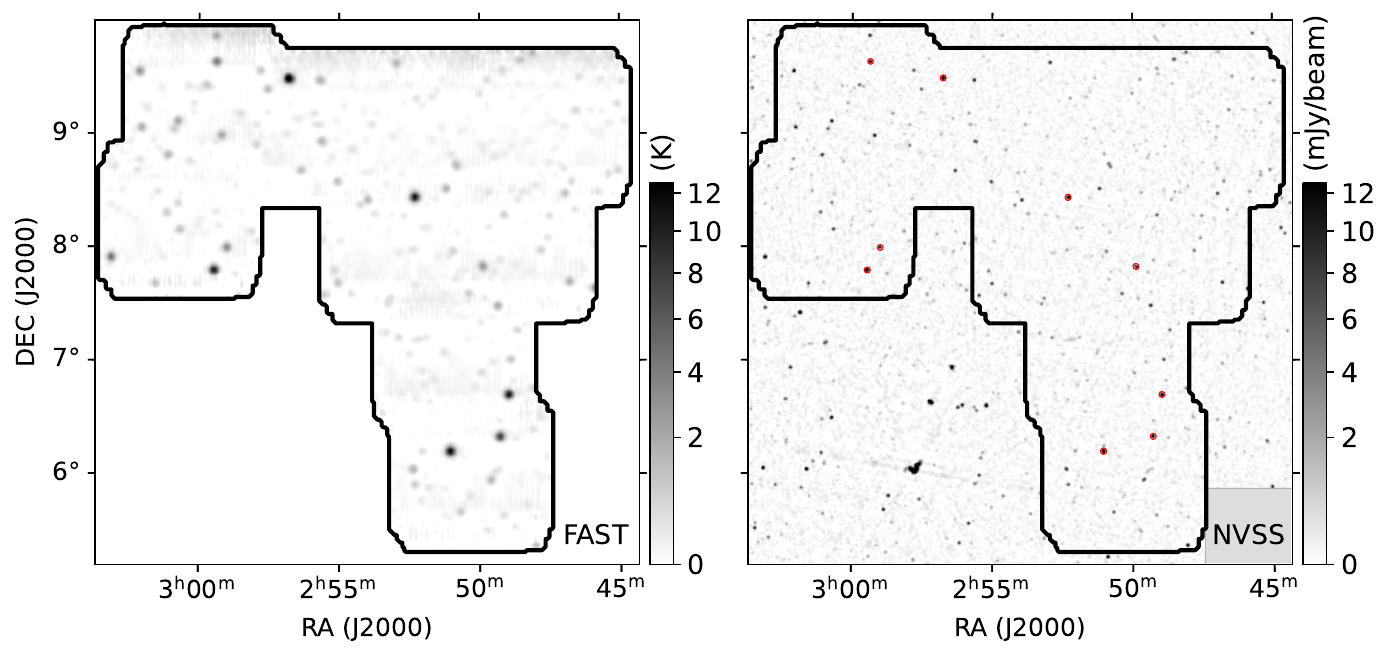}
\caption{
Comparison between the continuum emission of the G165 VHVC measured by FAST and the NVSS survey. 
\textbf{Left}: The 1.4 GHz continuum observed by FAST. 
\textbf{Right}: The 1.4 GHz continuum observed in the NVSS survey \citep{1998AJ....115.1693C}. 
The angular resolution of the right panel is 40$\arcsec$. 
The strong continuum sources observed by FAST ($T_{\rm peak} \gtrsim 3$ K) are fitted by 2D Gaussian functions and denoted by red circles in the right panel. 
The diameters of the red circles represent the FWHM of the fits.
\label{fig_conti}
}
\end{figure*}

\begin{figure*}[h]
\centering
\includegraphics[width=0.7\linewidth]{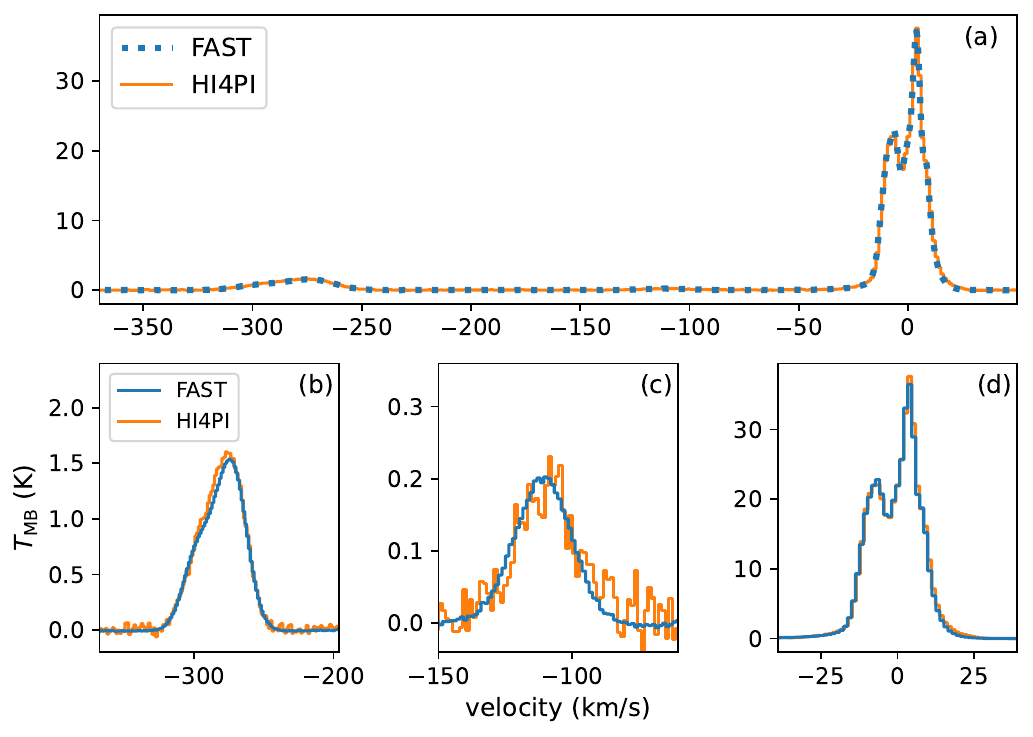}
\caption{
Comparison between the spectra of G165 VHVC from FAST and HI4PI. 
The spectra represent the mean spectra from a randomly chosen circle located at 2:52:55.6 +8:52:50.9 with a radius of 12$\arcmin$.
The FAST spectra have been spectrally smoothed and resampled to match the spectral resolution of the HI4PI data.
(\textbf{a}): The spectra over the full velocity range of H{\sc i} emission. The spectral baselines have been removed. 
(\textbf{b--d}): Zoom-ins to the velocity ranges of the VHVC (b), HVC (c), and foreground \HI~(d). 
The orange and blue lines in panels (b) and (d) are nearly identical. 
\label{fig_compare_spe}
}
\end{figure*}

\begin{figure*}[h]
\centering
\includegraphics[width=0.9\linewidth]{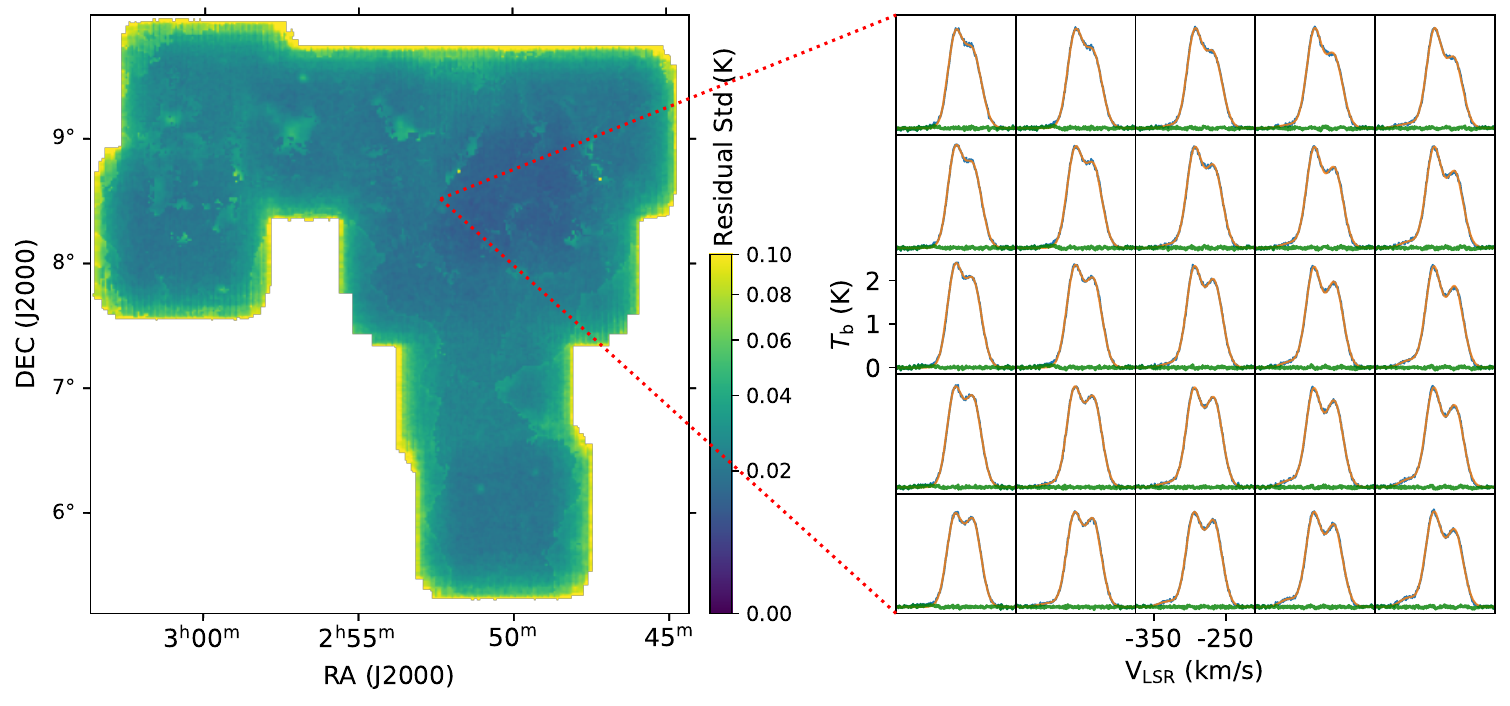}
\caption{The residuals of the multiple Gaussian fitting of G165 VHVC. 
\textbf{Left}: The standard deviation of the residual spectra.
\textbf{Right}: The blue lines represent the example spectra (Fig. 1), while the yellow lines correspond to the multiple Gaussian fitting results. The green lines indicates the residuals.
\label{fig_fit_residual}
}
\end{figure*}

\clearpage
\bibliographystyle{naturemag}

\end{document}